\documentclass[]{pasj02} 
\usepackage[switch,mathlines]{lineno} % add line number to manuscript
\usepackage{url}
\usepackage{amsmath}
\usepackage{arydshln}

\jyear{2025}
\Received{}%{yyyy/mm/dd}
\Accepted{}%{yyyy/mm/dd}
\newcommand{\bfp}{\mbox{\boldmath{$p$}}}

\newcommand{\bftheta}{\mbox{\boldmath{$\theta$}}}

\newcommand{\bfchi}{\mbox{\boldmath{$\chi$}}}

\begin{document} 

\title{Measurement of angular cross-correlation between the cosmological dispersion measure and the thermal Sunyaev--Zeldovich effect}
%Compton $y$ parameter}

%%% begin:list of authors
% Do NOT capitalize all letters in "textsc".
\author{
 Ryuichi \textsc{Takahashi},\altaffilmark{1} %\altemailmark\orcid{0000-0000-0000-0000} %\email{aaaaa@xxx.xxx.xx.xx} 
 Kunihito \textsc{Ioka},\altaffilmark{2}
 Masato \textsc{Shirasaki},\altaffilmark{3,4}
 and
 Ken \textsc{Osato}\altaffilmark{5--8}
}
\altaffiltext{1}{Faculty of Science and Technology, Hirosaki University, 3 Bunkyo-cho, Hirosaki, Aomori 036-8560, Japan}
\altaffiltext{2}{Yukawa Institute for Theoretical Physics, Kyoto University, Kyoto 606-8502, Japan}
\altaffiltext{3}{National Astronomical Observatory of Japan (NAOJ), National Institutes of Natural Sciences, Osawa, Mitaka, Tokyo 181-8588, Japan}
\altaffiltext{4}{The Institute of Statistical Mathematics, Tachikawa, Tokyo 190-8562, Japan}
\altaffiltext{5}{Center for Frontier Science, Chiba University, Chiba 263-8522, Japan}
\altaffiltext{6}{Department of Physics, Graduate School of Science, Chiba University, Chiba 263-8522, Japan}
\altaffiltext{7}{RIKEN Center for Advanced Intelligence Project, 1-4-1 Nihonbashi, Chuo, Tokyo 103-0027, Japan}
\altaffiltext{8}{Kavli Institute for the Physics and Mathematics of the Universe, The University of Tokyo Institutes for Advanced Study, Chiba,
277-8583, Japan}

%\footnotetext[$\dag$]{Present address: ....}

%%% end:list of authors

%% !!! Select 3 to 5 words from PASJ's key words !!! 
%% List of Key Words: https://academic.oup.com/pasj/pages/Pasj_Keywords 
%% "\KeyWords{ }" always has to be placed before ``\maketitle'' 

\maketitle

\begin{abstract}
The dispersion measures (${\rm DMs}$) from fast radio bursts (FRBs) and the thermal Sunyaev--Zeldovich (tSZ) effect probe the free-electron density and thermal pressure, respectively, in the intergalactic medium (IGM) and the intervening galaxies and clusters.
Their combination enables disentangling the gas density and temperature.
In this work, we present the first detection of an angular cross-correlation between the ${\rm DMs}$ and the Compton $y$ parameter of the tSZ effect.
The theoretical expectation is calculated using the halo model $\texttt{HMx}$, calibrated with hydrodynamic simulations.
The observational cross-correlation is measured over angular separations of  $1^\prime$--$1000^\prime$ using the ${\rm DMs}$ from $133$ localized FRBs and the $y$-maps from the Planck satellite and the Atacama Cosmology Telescope (ACT). 
We detect a positive correlation with amplitudes of $\mathcal{A}=2.01 \pm 0.50$ ($4.0 \sigma$) for Planck and $\mathcal{A}=1.23 \pm 0.82$ ($1.5 \sigma$) for ACT, where $\mathcal{A}=1$ corresponds to the theoretical prediction of the Planck 2018 $\Lambda$CDM cosmology. 
Assuming an isothermal gas, the measured amplitude implies an average electron temperature of $\approx 2 \times 10^7 \,  {\rm K}$.
The correlation is highly sensitive to the matter clustering parameter $\sigma_8$ and to baryon feedback, and its dependence on other cosmological and astrophysical parameters---such as the ionized fraction and the Hubble constant---differs from that of the ${\rm DM}$ alone. 
This suggests that future joint analyses of the ${\rm DMs}$ and the tSZ effect could help break degeneracies among these parameters.
\end{abstract}

%\pagewiselinenumbers 

\section{Introduction}

Fast radio bursts (FRBs) emit radio pulses, typically of several milliseconds in duration, across cosmological distances (e.g., \cite{Lorimer2007,Keane2016}). 
Although numerous theoretical models of FRBs have been proposed, their physical origin has not been consensually elucidated (e.g., \cite{Zhang2023}).
The dispersion measure (${\rm DM}$) of FRB, which measures the column density of free electrons along the line-of-sight to the source, can be determined from the frequency dependence of the pulse’s arrival time.
An FRB is called ``localized'' when its host galaxy has been identified and the redshift of that galaxy has been measured.
Currently, $\sim 130$ localized FRBs have been reported (summarized in Table \ref{table_list_FRBs1} of Appendix \ref{app:list_localized_FRB}), with a highest redshift of 2.15~\citep{Caleb2025}. 
The ${\rm DM}$ serves as a tool for exploring the cosmological distribution of free electrons, or equivalently, ionized gas.  

An angular auto-correlation of the ${\rm DM}$ has been proposed to measure the large-scale distribution of free electrons (e.g., \cite{Masui2015,Shirasaki2017,Reischke2021,TI2021,Saga2024}), but such an auto-correlation has not yet been detected~\citep{Xu2021}. 
Several theoretical studies have suggested cross-correlations between the ${\rm DM}$ and other signals such as foreground galaxies (e.g., \cite{McQuinn2014,Madhavacheril2019,Shirasaki2022,Sharma2025}), weak lensing~\citep{Reischke2023b}, and the thermal Sunyaev--Zeldovich (tSZ; \cite{SZ1970}) effect~\citep{MunozLoeb2018}. 
%The cross-correlation between the DM and galaxy weak lensing will constrain the baryon feedback strength~\citep{Reischke2023b}.
Because the observational data of galaxies, weak lensing, and tSZ are more abundant and higher quality than current ${\rm DM}$ data, the cross-correlation is expected to be more easily detected than the auto-correlation. 
Recently, an excess ${\rm DM}$ was observed around foreground galaxies~\citep{Connor2022,Wu2023,WangHaochen2025,Shirasaki2026} and filaments~\citep{Mo2025}.
Similarly, a cross-correlation with the number density of foreground galaxies was measured~\citep{Hsu2025,Hussaini2025}.

When cosmic microwave background (CMB) photons pass through a hot plasma, they gain energy from high-energy electrons via inverse Compton scattering, known as the tSZ effect (e.g., reviews by \cite{Kitayama2014,Mroczkowski2019}). 
The strength of the tSZ effect is characterized by the Compton $y$ parameter, which is proportional to the electron pressure integrated along the line-of-sight.
Multi-frequency CMB maps produce a $y$-map (e.g., \cite{Planck2016_ymap}).
The tSZ effect has been used to probe the gas properties (density, temperature, and entropy) of super-clusters (\cite{Tanimura2019b}), clusters (e.g., \cite{Planck2013}), galaxy halos (e.g., \cite{Planck2013}), and filaments between galaxies (e.g., \cite{deGraaff2019,Tanimura2019}).  
%and voids (e.g.,~\cite{Alonso2018}).
Angular cross-correlations between the $y$ parameter and the weak lensing signal have already constrained cluster physics (e.g., density and pressure profiles and the hydrostatic mass bias) and cosmology (e.g., \cite{VanWaerbeke2014,Ma2015,Osato2018,Osato2020,LaPosta2024,Pandey2025}). 

In this paper, we measure the cross-correlation between the $y$ parameter and the cosmological component of ${\rm DM}$ (denoted as ${\rm DM}_{\rm cos}$) caused by ionized gas in the intergalactic medium (IGM) and in intervening galaxies and clusters. 
To our knowledge, this is the first measurement of the $y$-${\rm DM}_{\rm cos}$ correlation.
%The DM probes the free-electron density and the tSZ probes its pressure (its density times its temperature); therefore, the coss-correlation can provide the temperature~\citep{MunozLoeb2018}.
\citet{Fujita2017} demonstrated that if sufficiently many FRB events occur behind a nearby massive cluster, the electron number density and temperature profiles of that cluster can be determined by combining the FRB ${\rm DMs}$ and the $y$-map.
\citet{Connor2023} found two localized FRBs with host clusters and estimated the gas temperatures of the intracluster medium (ICM) by combining the ${\rm DMs}$ and $y$.
\citet{MunozLoeb2018} theoretically studied a $y$-${\rm DM}$ correlation to estimate the number of FRBs required for extracting the temperature of the warm–hot intergalactic medium (WHIM) from the correlation signal.
Our research differs from \citet{MunozLoeb2018} in the following ways: 1) Whereas they considered the $y$ and ${\rm DM}$ at the same sky position, we correlate them within an angular separation of less than $1000^\prime$, significantly increasing the number of correlation pairs and enhancing the resulting signal-to-noise ratio.
2) We account for the spatial fluctuations of free-electron density, which they did not consider. 
As a result, their correlation mainly arises from differences in source redshift (i.e., higher/lower ${\rm DMs}$ for distant/nearby sources). 
3) We utilize localized FRBs, which offer several advantages over the unlocalized FRBs considered in their study.
First, since the redshifts are known, the average extragalactic ${\rm DM}$ (denoted as ${\rm  DM}_{\rm ext}$) at a given redshift can be estimated from the ${\rm DM}$--$z$ relation (e.g., \cite{Palmer1993,Ioka2003,Inoue2004,DZ2014}). 
The residual from the average ${\rm DM}_{\rm ext}$ traces the fluctuations in free-electron density. 
Second, the angular positions of localized FRBs can be determined much more accurately (to sub-arcsecond scales) than those of unlocalized FRBs ($\sim 0.2 \, {\rm deg}$ for the Canadian Hydrogen Intensity Mapping Experiment (CHIME)\footnote{\url{https://www.chime-frb.ca/catalog}}), enabling smaller-scale correlation measurements.
Although the smaller number of localized FRBs compared to unlocalized FRBs is a current disadvantage, localized FRB events are being quickly accumulated thanks to ongoing detectors such as the CHIME/FRB outriggers~\citep{CHIME_Outriggers2025}, the Deep Synoptic Array (DSA)\footnote{\url{https://www.deepsynoptic.org}}, and the Commensal Real-time ASKAP Fast Transients survey (CRAFT)\footnote{\url{https://research.curtin.edu.au/cira/our-research/science/craft-survey/}}.

The remainder of this paper is organized as follows. 
We first derive a theoretical angular cross-correlation between ${\rm DM}_{\rm cos}$ and $y$ using the halo model \texttt{HMx} (\cite{Mead2020}; Section \ref{sec:theory}). 
A simple phenomenological model assuming constant gas temperature is also introduced (Subsection \ref{sec:pk_constTe}). 
Section \ref{sec:host_galaxy_corr} estimates the correlation between the host ${\rm DM}$ and $y$ based on \texttt{HMx}, which might contaminate the cosmological correlation signal. 
Section \ref{sec:obs_data} describes our observational data: Subsection \ref{sec:full-skymap_DeltaDMe} calculates the average ${\rm DM}_{\rm ext}$ from $133$ localized FRBs using the ${\rm DM}$--$z$ relation and then derives the ${\rm DM}_{\rm ext}$ residual by subtracting the average. 
Subsection \ref{ssec:ymap} presents the $y$-maps from Planck and ACT. 
Section \ref{sec:measurment} introduces an estimator of the cross-correlation between the ${\rm DM}_{\rm ext}$ residual and $y$ (Subsection \ref{sec:w_yDM_estimator}) and presents our main measurement results along with the theoretical predictions (Subsection \ref{sec:w_yDM_results}). 
We also constrain the gas temperature based on the correlation amplitude (Subsection \ref{sec:constraint_Te}). 
Section \ref{sec:discussion} discusses potential contamination in the cross-correlation measurement, and Section \ref{sec:conclusion} concludes the paper.

This paper assumes a spatially flat $\Lambda$CDM model consistent with the Planck 2018 best-fitting parameters~\citep{Planck2020}: matter density $\Omega_{\rm m}=1-\Omega_\Lambda=0.315$, baryon density $\Omega_{\rm b}=0.049$, Hubble parameter $h=0.674$, spectral index $n_{\rm s}=0.965$, and amplitude of matter density fluctuations on the scale of $8 \, h^{-1}$Mpc $\sigma_8=0.811$.
Except for the gas temperature, all physical quantities such as length, wavenumber, number density, and pressure are expressed in comoving units.

\section{Theoretical model of angular cross-correlation of ${\rm DM}_{\rm cos}$ and $y$} 
\label{sec:theory}

The observed ${\rm DM}$ is decomposed into its Milky Way (MW), cosmological, and host contributions as follows:
\begin{equation}
    {\rm DM}_{\rm obs} =  {\rm DM}_{\rm MW} +{\rm DM}_{\rm cos} + {\rm DM}_{\rm host}.
\label{eq:DM_obs}
\end{equation}
Here, ${\rm DM}_{\rm cos}$ includes contributions from the IGM, intervening galaxies and clusters. 
${\rm DM}_{\rm host}$ is contributed by the host galaxy, including the host cluster if it is part of the galaxy cluster, as seen in recently discovered FRBs~\citep{Connor2023,Amiri2025}.  
Meanwhile, ${\rm DM}_{\rm MW}$ can be inferred from models of the free-electron distribution in the Galactic interstellar medium (ISM) and halo. 
We utilize the NE2001~\citep{NE2001} or YMW16~\citep{YMW2017} model\footnote{\texttt{PyGEDM}~\citep{Price2021} is used.} for the ISM and the YT20 model~\citep{YT2020} for the halo.
The extragalactic contribution is then obtained as 
\begin{align}
     {\rm DM}_{\rm ext} &\equiv  {\rm DM}_{\rm obs} -{\rm DM}_{\rm MW},  \nonumber \\
                          &= {\rm DM}_{\rm cos} + {\rm DM}_{\rm host}.
     \label{eq:DM_ext}
\end{align}
To examine the correlation between ${\rm DM}_{\rm ext}$ and $y$, the remainder of this section computes the correlation between ${\rm DM}_{\rm cos}$ and $y$ and Section \ref{sec:host_galaxy_corr} discusses the correlation between ${\rm DM}_{\rm host}$ and $y$.

The hot gas in the MW can create a correlation between ${\rm DM}_{\rm MW}$ and $y$. 
If the electron density model of MW is sufficiently accurate, ${\rm DM}_{\rm ext}$ and $y$ are not correlated within MW (because ${\rm DM}_{\rm ext}$ excludes the MW contribution). 
As the electron density model contains uncertainties, the error in ${\rm DM}_{\rm MW}$ could introduce an additional correlation between ${\rm DM}_{\rm ext}$ and $y$, which is ignored in the paper.

\subsection{The cosmological ${\rm DM}$}
\label{sec:cosmo_DM}

We consider the ${\rm DM}_{\rm cos}$ of an FRB at angular position $\bftheta$ in the sky with redshift $z_{\rm s}$.
A free-electron gas lies along the line-of-sight to the FRB at spatial position $\bfchi$ and redshift $z$.
Here, $\bfchi$ points from the observer to the source, and its absolute value is the comoving distance: $\chi(z)=c \int_0^z {\rm d}z^\prime /H(z^\prime)$ where $H(z)$ is the Hubble expansion rate.
The vector $\bfchi$ can be decomposed into radial ($\chi$) and two-dimensional perpendicular ($\chi \bftheta$) components. % with $|\bftheta| \ll 1$. %: $\bfr=(r,r \bftheta)$
The ${\rm DM}_{\rm cos}$ is the column density of free electrons along the line of sight (e.g., \cite{Ioka2003,Inoue2004}):
\begin{equation}
    {\rm DM}_{\rm cos}(\bftheta;z_{\rm s}) = \int_0^{z_{\rm s}} \!\! \frac{c \, {\rm d}z}{H(z)} n_{\rm e}(\bfchi;z) (1+z).
\label{eq:DM}
\end{equation}
The free-electron density can be decomposed into its spatial mean and fluctuations: 
\begin{equation}
    n_{\rm e}(\bfchi;z_{\rm s}) = \bar{n}_{\rm e}(z) \left[ 1 + \delta_{\rm e}(\bfchi;z) \right].
\label{eq:dne}
\end{equation}
The spatial average of the second term vanishes, i.e., $\langle \delta_{\rm e} \rangle=0$.
The mean free-electron density is~(e.g., \cite{DZ2014})
\begin{equation}
    \bar{n}_{\rm e}(z) = \frac{3 H_0^2}{8 \pi G} \frac{\Omega_{\rm b}}{m_{\rm p}} f_{\rm e}(z) \left( X(z) + \frac{1}{2} Y(z) \right),
\label{eq:ne_mean}
\end{equation}
where $m_{\rm p}$ is the proton mass, and $X$ and $Y$ are the mass fractions of hydrogen and helium, respectively, here set to $X=1-Y=0.75$.
Based on the ${\rm DM}$--$z$ relation with localized FRBs, the ionized fraction $f_{\rm e}$ is currently constrained to $f_{\rm e} \approx 0.8$--$1$ (e.g., \cite{Li2020,Lemos2023,Wang2023,Khrykin2024,Connor2024}).   
As the redshift evolution of $f_{\rm e}$ has not been well constrained (e.g., \cite{Lin2023,Liu2025}), we assume that $f_{\rm e}$ is constant. 
Substituting Eq.~(\ref{eq:dne}) into Eq.~(\ref{eq:DM}), the cosmological ${\rm DM}$ is separated into an isotropic component ($\overline{\rm DM}_{\rm cos}$) and angular fluctuations ($\delta {\rm DM}_{\rm cos}$):
\begin{align}
  \overline{\rm DM}_{\rm cos}(z_{\rm s}) &= \int_0^{z_{\rm s}} \!\! \frac{c \, {\rm d}z}{H(z)} \bar{n}_{\rm e}(z) (1+z), \nonumber \\
  \delta {\rm DM}_{\rm cos}(\bftheta;z_{\rm s}) &= \int_0^{z_{\rm s}} \!\! \frac{c \, {\rm d}z}{H(z)} \bar{n}_{\rm e}(z) \delta_{\rm e} (\bfchi;z) (1+z).
\label{eq:DMcos}
\end{align}

The variance of the fluctuations is given by (e.g., \cite{McQuinn2014}),
\begin{align}
    \sigma_{\rm DM,cos}^2(z_{\rm s}) &\equiv \langle \left[ \delta {\rm DM}_{\rm cos}(\bftheta;z_{\rm s}) \right]^2 \rangle \nonumber \\
    &= \int_0^{z_{\rm s}} \!\! \frac{c \, {\rm d}z}{H(z)} \bar{n}_{\rm e}^2(z) (1+z)^2 \int_0^\infty \!\! \frac{k\,{\rm d}k}{2 \pi}  P_{n_{\rm e}}(k;z),
\label{eq:sigma_DMcos}
\end{align}
where $P_{n_{\rm e}}(k;z)$ is the power spectrum of the free-electron density contrast as a function of wavenumber ($k$) and redshift ($z$).

\subsection{The Compton $y$ parameter}

When CMB photons pass through a hot gas, thermal electrons transfer their kinetic energy to the photons via inverse Compton scattering.
The resulting distortion of the CMB spectrum is known as the tSZ effect.
The strength of the distortion is specified by the dimensionless Compton $y$ parameter, obtained by integrating the electron pressure along the line-of-sight: 
\begin{equation}
    y(\bftheta) = \frac{\sigma_{\rm T}}{m_{\rm e} c^2} \int_0^{z_*} \!\! \frac{c \, {\rm d}z}{H(z)} p_{\rm e}(\bfchi;z) (1+z)^2,
\label{eq:y}
\end{equation}
where $z_*$ is the redshift of the last scattering surface, $\sigma_{\rm T}$ is the Thomson-scattering cross section, and $m_{\rm e}$ is the electron mass, and $p_{\rm e}$ is the electron pressure in the comoving unit (the physical quantity is $p_{\rm e} (1+z)^3$). % of ${\rm eV}/{\rm cm}^3$.
For an ideal gas, $p_{\rm e}$ is related to the physical temperature $T_{\rm e}$ as $p_{\rm e}=n_{e} k_{\rm B} T_{\rm e}$, where $k_{\rm B}$ is the Boltzmann constant.
As the pressure is decomposed into its spatial mean and fluctuations, $p_{\rm e}(\bfchi;z)=\bar{p}_{\rm e}(z)+\delta p_{\rm e}(\bfchi;z)$, the $y$ parameter is similarly decomposed as
\begin{equation}
    y(\bftheta) = \bar{y} + \delta y(\bftheta).
\end{equation}

Because the $y$ parameter measured by Planck and ACT is smeared over the finite beam size of the detector, we apply a smoothing filter to $y$:
\begin{equation}
    y_{\rm sm}(\bftheta)=\int \! {\rm d}^2\bftheta^\prime \, W_{\rm sm} (\bftheta-\bftheta^\prime) \, y(\bftheta^\prime),
\label{eq:y_smooth}
\end{equation}
where $W_{\rm sm}$ is the smoothing kernel. 
In the absence of smoothing, $W_{\rm sm}$ is replaced with the Dirac delta function; $W_{\rm sm}=\delta_{\rm D}^2 (\bftheta-\bftheta^\prime)$.

\subsection{Angular cross-correlation of $y$ and ${\rm DM}_{\rm cos}$}

Under the Limber and flat-sky approximations (e.g., \cite{BS2001}), the angular cross-correlation of $y_{\rm sm}(\bftheta_1)$ and ${\rm DM}_{\rm cos}(\bftheta_2)$ at separation $\theta$ ($=|\bftheta_1-\bftheta_2|$) is written as
\begin{align}
  w_{y {\rm DM}}^{\rm (theo)}(\theta;z_{\rm s}) &\equiv \langle  \delta y_{\rm sm}(\bftheta_1) \, \delta {\rm DM}_{\rm cos} (\bftheta_2;z_{\rm s})\rangle \nonumber \\
  & = \frac{\sigma_{\rm T}}{m_{\rm e} c^2} \int_0^{z_{\rm s}} \!\! \frac{c \, {\rm d}z}{H(z)} (1+z)^3 \bar{n}_{\rm e}(z) \int_0^\infty \!\!\frac{k \, {\rm d}k}{2 \pi} \nonumber \\
  & ~\times P_{n_{\rm e}p_{\rm e}}(k;z) \int \! {\rm d}^2\bftheta^\prime \, W_{\rm sm} (\bftheta-\bftheta^\prime) J_0(k\chi(z)\theta^\prime),
\label{eq:w_yDM}
\end{align}
where $J_0$ is the zero-th order Bessel function and $P_{n_{\rm e}p_{\rm e}}(k;z)$ is the cross-power spectrum of the electron density contrast ($\delta_{\rm e}$) and pressure fluctuations ($\delta p_{\rm e}$).
Equation (\ref{eq:w_yDM}) is valid for small angular separations ($\theta \ll 1 \, {\rm rad}$) under the flat-sky approximation.
As the electron fraction is proportional to $f_{\rm e}$, the overall amplitude of $w_{y {\rm DM}}^{\rm (theo)}$ scales with $f_{\rm e}^2$.

\subsection{Halo model \texttt{HMx}}
\label{sec:HMx}

This subsection presents a theoretical model of the power spectra $P_{n_{\rm e}}$ and $P_{n_{\rm e}p_{\rm e}}$  (Subsubsection \ref{sec:halomodel_HMx}) and the resulting cross-correlation $w_{y{\rm DM}}^{\rm (theo)}$ (Subsubsection \ref{sec:theoretical_w_yDM}) based on the halo model \texttt{HMx}.

\subsubsection{\texttt{HMx}}
\label{sec:halomodel_HMx}

We use the public code\footnote{The source code library of Fortran90 functions in \url{https://github.com/alexander-mead/library}.} \texttt{HMx}~\citep{Mead2020,Troster2022} to obtain $P_{n_{\rm e}}$ and $P_{n_{\rm e}p_{\rm e}}$. %these power spectra. 
\texttt{HMx} utilizes the halo model framework (e.g., \cite{CS2002,Arico2020,Shirasaki2022,Asgari2023}), in which the model components (including the gas and temperature profiles within a halo) were calibrated through hydrodynamic simulations BAryons and HAloes of MAssive Systems (BAHAMAS)~\citep{McCarthy2017,McCarthy2018}.
We employ model (3) in Table 2 of \citet{Mead2020}.
The model parameters were determined to reproduce the auto- and cross-power spectra of total matter and electron pressure measured in BAHAMAS.
The calibration range is $k=0.015$--$7 \, h {\rm Mpc}^{-1}$ and $z=0$--$1$.
\texttt{HMx} includes three mass components: CDM, gas, and stars. 
The gas is assumed to be fully ionized with all free electrons included.
$P_{n_{\rm e}}$ is obtained from the auto-power spectrum of gas density in \texttt{HMx} denoted as $P_{\rm gas}^{\rm HMx}$, assuming that free electrons exactly trace the gas (i.e., $\delta_{\rm e}=\delta \rho_{\rm gas}/\bar{\rho}_{\rm gas}$).
Similarly, $P_{n_{\rm e}p_{\rm e}}$ is obtained from the cross-power spectrum of gas and electron pressure\footnote{\texttt{HMx} uses the electron thermal pressure (defined as the product of the electron number density and temperature), tailored for the tSZ effect. Therefore, in our understanding, the pressure does not include the non-thermal component.} $P_{{\rm gas},p_{\rm e}}^{\rm HMx}$ in \texttt{HMx}. 
Because the normalizations of density perturbations ($\delta_{\rm e}$ and $\delta p_{\rm e}$) in \texttt{HMx} differ from ours, we rescale them as shown in Appendix \ref{app:pk_norm} (see also \cite{RT2024}).

Baryonic feedback expels a fraction of the gas within a halo to the outside, dividing it into bound and ejected components.
The feedback strength of an active galactic nucleus (AGN) is determined by the heating temperature $T_{\rm AGN}$, defined as the temperature increase of the gas particles targeted for feedback.
\texttt{HMx} was calibrated at three temperatures:  $\log_{10}(T_{\rm AGN}/{\rm K}) = 7.6$, $7.8$, and $8.0$, where $7.8$ is the fiducial value used to reproduce the observed hot gas fraction in groups and clusters~\citep{McCarthy2017}.
\texttt{HMx} describes the density and temperature profiles of the bound gas embedded in a CDM halo~\citep{NFW1997} using the \citet{KS2001} model (also \cite{Martizzi2013}).
The electron number density and pressure profiles in a halo are discussed in Subsection 3.3 of \citet{Mead2020}.
The ejected gas traces the linear matter density field, which has a temperature of $\approx 10^{6.5} \, {\rm K}$ (suggested as the WHIM temperature based on cross-correlation measurement between the tSZ signal and weak lensing; \cite{VanWaerbeke2014}).  %, which contributes only to the 2-halo term. 

The power spectrum is decomposed into 1- and 2-halo terms:
\begin{equation}
    P(k;z) =  P^{\rm 1h}(k;z) + P^{\rm 2h}(k;z),
\end{equation}
where $P$ denotes $P_{n_{\rm e}}$ or $P_{n_{\rm e} p_{\rm e}}$.
The first term arises from a correlation within the same halo, which dominates on small scales ($k \gtrsim$ some $h \, {\rm Mpc}^{-1}$ at $z=0$--$1$), while the second term stems from a correlation between two different halos and the ejected (diffused) gas, which dominates on large scales.

\subsubsection{Theoretical cross-correlation results}
\label{sec:theoretical_w_yDM}

\begin{figure}
 \begin{center}
   \hspace*{-0.5cm}
   \includegraphics[width=10.5cm]{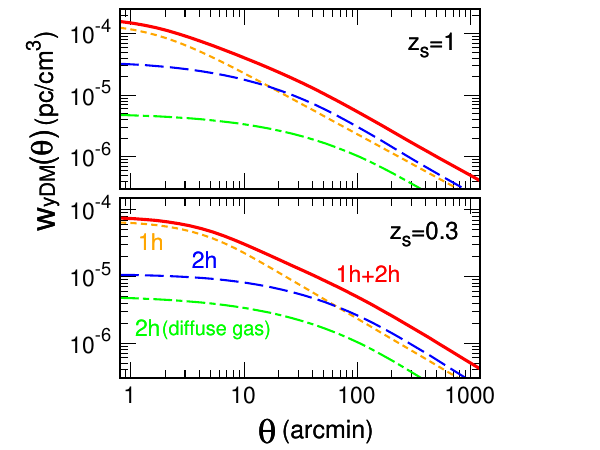}  %{fig_xi_paper.pdf} 
 \end{center}
\caption{Theoretical angular cross-correlation of $y$ and ${\rm DM}_{\rm cos}$ for $z_{\rm s}=1$ (top) and $0.3$ (bottom) obtained with \texttt{HMx}. The dotted orange and dashed blue curves represent the 1- and 2-halo terms, respectively, and the solid red curve is their sum. The dot-dashed green curve indicates the diffuse gas contribution in the 2-halo term. Here, we assume $f_{\rm e}=0.9$, and the overall amplitudes scale proportionally to $f_{\rm e}^2$. 
%{Alt text: Line graphs showing that the correlation decreases as angular separation increases.}
}
\label{fig_xi}
\end{figure}

\begin{figure*}
 \begin{center}
   \includegraphics[width=17cm]{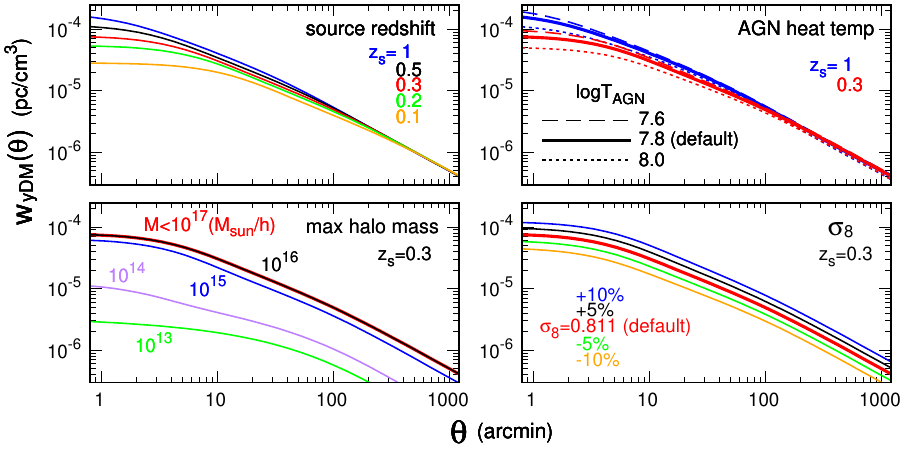}  %{fig_xi_param-depend_paper.pdf} 
 \end{center}
\caption{Similar to Fig.~\ref{fig_xi}, but plotting the parameter dependencies of $w_{y{\rm DM}}^{\rm (theo)}(\theta;z_{\rm s})$. The top-left panel plots results at different source redshifts: $z_{\rm s}=1, 0.5, 0.3, 0.2$ and $0.1$ from top to bottom. 
The top-right panel displays results for different AGN heating temperatures at $z_{\rm s}=0.3$ and $1$: $\log_{10} (T_{\rm AGN}/{\rm K})=7.6, 7.8$, and $8.0$.
The bottom-left panel presents results for different maximum halo virial masses at $z_{\rm s}=0.3$. The thick red curve corresponds to the default mass range $10^7 < M/(h^{-1} M_\odot) < 10^{17}$, and the other curves alter the maximum mass to $10^{16}, 10^{15}, 10^{14}$, and $10^{13} \, h^{-1} M_\odot$ from top to bottom. The red and black curves overlap.  The bottom-right panel shows results for various $\sigma_8$ at $z_{\rm s}=0.3$. The thick red curve represents the default, and the other curves change the default value by $10 \%$, $5 \%$, $-5 \%$, and $-10 \%$ from top to bottom. In all panels, $f_{\rm e}=0.9$.
%{Alt text: Line graphs illustrating that the correlation strongly depends on the maximum halo mass, sigma 8, and $T_{\rm AGN}$ (at small scales)}.
}
\label{fig_xi_param}
\end{figure*}

Figure \ref{fig_xi} plots the angular cross-correlation $w_{y{\rm DM}}^{\rm (theo)}(\theta;z_{\rm s})$ obtained using Eq.~(\ref{eq:w_yDM}) with $f_{\rm e}=0.9$, $\log_{10}(T_{\rm AGN}/{\rm K})=7.8$, and no smoothing on $y$. 
The result decreases approximately proportionally to $\theta^{-1}$ for $\theta \gtrsim 10^\prime$.
The 1- and 2-halo terms are comparable at $\theta \sim 65^\prime$ for $z_{\rm s}=0.3$ and at $\theta \sim 17^\prime$ for $z_{\rm s}=1$.
The larger contribution of diffuse gas at lower redshifts than at higher redshifts can be attributed to baryon feedback, which increases the abundance of diffuse gas.
The remaining 2-halo term results from the correlation of gas between separate halos.

Figure \ref{fig_xi_param} illustrates the dependencies of the cross-correlation $w_{y{\rm DM}}^{\rm (theo)}(\theta;z_{\rm s})$ on the model parameters.
As shown in the top-left panel, smaller-scale amplitudes are more sensitive to the source redshift.
This behavior can be attributed to two phenomena.
First, the 1-halo term is primarily determined by an abundance of massive halos ($M \gtrsim 10^{14} \, h^{-1} M_\odot$), which are sensitive to redshift.
Second, smaller (larger) scale signals are primarily affected by distant (nearby) structures owing to their apparent angular size.
The redshifts of the 133 localized FRBs (listed in Appendix \ref{app:list_localized_FRB}) range from $0.004$ to $2.15$, with an average of $0.26$.
As shown in the top-right panel, the signal is suppressed at higher $T_{\rm AGN}$, especially in the 1-halo term, because the gas in halos is more effectively expelled at higher $T_{\rm AGN}$; therefore, the small-scale signal is sensitive to $T_{\rm AGN}$.
Hereafter, $T_{\rm AGN}$ is set to $\log_{10} (T_{\rm AGN}/{\rm K})=7.8$ unless stated otherwise.
The bottom-left panel shows the halo mass dependence, where the mass includes the diffuse gas ejected from halos through feedback (Subsection 2.2 of \cite{Mead2020}).  
The signal is predominantly contributed by massive halos of $M \gtrsim 10^{14} \, h^{-1}M_\odot$, which contain a large amount of hot gas.
This halo-mass dependence is the same as that for the tSZ angular power spectrum (e.g., \cite{KS2002}). 
The bottom-right panel shows that the cross-correlation amplitude varies with $\sigma_8$ on both small and large scales because the abundance of massive halos is highly sensitive to $\sigma_8$.
%The default value is $\log(T_{\rm W}/{\rm K}) \approx 6.5$, which decreases slightly with redshift ($\log(T_{\rm W}/{\rm K}) = 6.65$ at $z=0$ and $6.50$ at $z=1$).
%The default $T_{\rm W}$ is consistent with the cross-correlation result between $y$ and weak-lensing convergence (e.g., \cite{VanWaerbeke2014}).

Let us now examine the input-parameter dependence of the cross-correlation $w_{y{\rm DM}}^{\rm (theo)}(\theta;z_{\rm s})$.
In addition to the parameters $\log_{10} (T_{\rm AGN}/K)$ and $\sigma_8$ (plotted in Fig.~\ref{fig_xi_param}), we vary $h$ and $\Omega_{\rm m}$ by $\pm 5\%$ around the fiducial cosmological model (while $\Omega_{\rm b}$ and $\Omega_{\rm m}+\Omega_\Lambda$ remain fixed) to compute finite differences of $w_{y{\rm DM}}^{\rm (theo)}$ with respect to these parameters. 
The function $w_{y{\rm DM}}^{\rm (theo)}(\theta;z_{\rm s})$ approximately depends on these parameters as follows:
\begin{align}
    & w_{y{\rm DM}}^{\rm (theo)}(\theta;z_{\rm s}) \nonumber \\
    & ~~~\propto f_{\rm e}^2 \, h^{1.7} \, \Omega_{\rm m}^{0.3} \, \sigma_8^{5.4} \, [\log_{10} (T_{\rm AGN}/{\rm K})]^{-6.5} 
    ~~{\rm for}~\theta=10^\prime, \nonumber \\
    & ~~~\propto f_{\rm e}^2 \, h^{1.5} \, \Omega_{\rm m}^{0.0} \, \sigma_8^{4.8} \, [\log_{10} (T_{\rm AGN}/{\rm K})]^{-3.4} 
    ~~{\rm for}~\theta=100^\prime,
\label{eq:w_theo_param-depend}
\end{align}
at $z_{\rm s}=0.3$.
Note that the cross-correlation is quite sensitive to $\sigma_8$, similar to the tSZ power spectrum (e.g., \cite{KS2002}), and to $T_{\rm AGN}$ on small scales.
The dependence on $h$ partially arises from ${\rm DM}_{\rm cos} \propto h$.
The parameter dependence in Eq.~(\ref{eq:w_theo_param-depend}) differs from that of ${\rm DM}_{\rm cos}$ ($\propto f_{\rm e} h$) in the ${\rm DM}$--$z$ relation; therefore, combining these probes can strongly constrain these parameters by breaking the parameter degeneracy. 

\subsection{A constant gas temperature model}
\label{sec:pk_constTe}

\begin{figure}
 \begin{center}
   \includegraphics[width=8.cm]{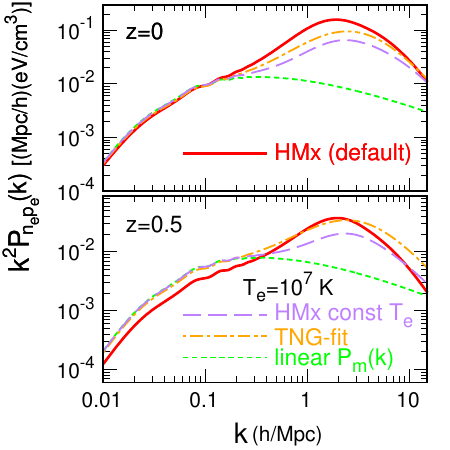}  %{fig_pk_paper.pdf} 
 \end{center}
\caption{Cross-power spectra of free-electron density and pressure (in units of ${\rm eV}/{\rm cm}^3$) at $z=0$ (top) and $0.5$ (bottom).
The solid red curve represents our default \texttt{HMx} model (Subsection \ref{sec:HMx}). 
The other curves correspond to the constant gas-temperature model with $T_{\rm e}=10^7 {\rm K}$, using different $P_{n_{\rm e}}$ models: \texttt{HMx} (dashed purple curve), a fitting function from the TNG300 simulation (dot-dashed orange curve), and the linear matter power spectrum (dotted green curve).
The amplitudes of these three curves scale as $\propto (f_{\rm e}/0.9)^2 (T_{\rm e}/10^7 {\rm K})$. 
%{Alt text: The power spectra on the $y$-axis are multiplied by the wavenumber squared for clearer presentation.}
}
\label{fig_pk}
\end{figure}

\begin{figure}
 \begin{center}
   \hspace*{-0.5cm}
   \includegraphics[width=10.5cm]{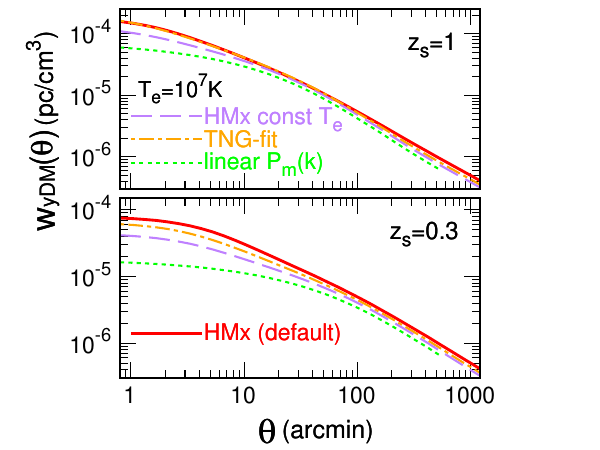}  %{fig_xi_constTe_paper.pdf} 
 \end{center}
\caption{Similar to Fig. \ref{fig_xi}, but plotting $w_{y{\rm DM}}^{\rm (theo)}(\theta;z_{\rm s})$ for the constant gas-temperature model with $T_{\rm e}=10^7 {\rm K}$. The curves are described in the captions of Fig.~\ref{fig_pk}. 
%{Alt text: Line graphs.}
}
\label{fig_xi_constTe}
\end{figure}

We also examine a simple phenomenological model assuming a constant gas temperature $T_{\rm e}$.
Using the equation of state ($p_{\rm e}=n_{\rm e} k_{\rm B} T_{\rm e}$), the cross-power spectrum is rewritten as:
\begin{equation}
  P_{n_{\rm e}p_{\rm e}}(k;z) = \bar{n}_{\rm e}(z) k_{\rm B} T_{\rm e} P_{n_{\rm e}}(k;z),
\label{eq:P_nepe_constTe}
\end{equation}
from which $P_{n_{\rm e} p_{\rm e}}$ can be obtained for a given $T_{\rm e}$ and $P_{n_{\rm e}}$.
To obtain $P_{n_{\rm e}}$, alongside \texttt{HMx}, we also consider a fitting function calibrated with the simulation suit IllustrisTNG300 (e.g., \cite{Springel2018,Nelson2019})\footnote{\url{https://www.tng-project.org}}.
Hereafter, this model will be referred to as TNG-fit.
As \texttt{HMx} and TNG-fit were calibrated through different hydrodynamic simulations, they can be compared for examining baryon feedback effects on the cross-correlation.
In TNG-fit, $P_{n_{\rm e}}$ is written as 
\begin{equation}
    P_{n_{\rm e}}(k;z) = b^2_{\rm e}(k;z) P_{\rm DMO}(k;z),
\end{equation}
where $b_{\rm e}$ is a fitting function of the free-electron bias (\cite{TI2021}) and $P_{\rm DMO}$ is the non-linear matter power spectrum in dark-matter-only (DMO) simulations, obtained using \texttt{halofit} (\cite{Smith2003,RT2012}).

Figure \ref{fig_pk} plots the cross-power spectrum $P_{n_{\rm e}p_{\rm e}}$ at $z=0$ and $0.5$ derived from our default HMx model (Subsection \ref{sec:HMx}) and from the constant $T_{\rm e}$ model.
The linear matter power spectrum gives a lower amplitude than the others at $k \gtrsim 0.1 \, h/{\rm Mpc}$ because it does not account for non-linear gravitational evolution.
The three constant $T_{\rm e}$ models converge at large scales because the free-electron distribution follows the underlying matter distribution at those scales (i.e., $b_{\rm e} \simeq 1$; \cite{TI2021}).
%According to \texttt{HMx} (where $P_{n_{\rm e}p_{\rm e}} \propto T_{\rm e}$ from Eq.~(\ref{eq:P_nepe_constTe})), 
From the large-scale amplitudes of the four curves, the corresponding gas temperature on large scales is $T_{\rm e} \simeq 10^7 \, {\rm K}$ at $z = 0$ and $T_{\rm e} \simeq 5 \times 10^6 \, {\rm K}$ at $z = 0.5$ (because $P_{n_{\rm e}p_{\rm e}} \propto T_{\rm e}$ from Eq.~(\ref{eq:P_nepe_constTe})).
The higher temperature at $z = 0$ than at $z = 0.5$ is explained by the formation of massive halos.
At $z=0$, the solid red and dashed purple curves, both derived from \texttt{HMx}, agree on large scales, but the red curve exceeds the purple one on small scales. 
This discrepancy arises because the gas temperature increases inside matter clumps such as groups and clusters.
The TNG-fit produces the largest signal among the constant $T_{\rm e}$ models because baryon feedback in TNG300 is weaker than in BAHAMAS with $\log_{10} (T_{\rm AGN}/{\rm K})=7.8$ (e.g., \cite{Chisari2019}).

Figure \ref{fig_xi_constTe} plots the cross-correlation $w_{y {\rm DM}}^{\rm (theo)}(\theta;z_{\rm s})$ obtained by inserting Eq.~(\ref{eq:P_nepe_constTe}) into Eq.~(\ref{eq:w_yDM}).
The constant $T_{\rm e}$ models with $T_{\rm e}=10^7 \, {\rm K}$ and the default \texttt{HMx} model produce similar amplitudes on large scales ($\theta > 10^\prime$) because the free-electron distribution at lower redshifts contributes to the cross-correlation on larger scales (as shown in the top-left panel of Fig.~\ref{fig_xi_param}).
The TNG-fit predicts a slightly higher amplitude on large scales ($\theta \gtrsim 100^\prime$) than the other constant $T_{\rm e}$ models because the large-scale correlations are partly contributed by small-scale $P_{n_{\rm e}p_{\rm e}}$ at low redshifts.

\section{Host contribution to the cross-correlation}
\label{sec:host_galaxy_corr}

This section estimates the host-induced cross-correlation with HMx. 
Subsection \ref{sec:yDM_host} computes the contribution of each halo mass to the cross-correlation, Subsection \ref{sec:yDM_host_ave} evaluates the ensemble average over halo masses, and Subsection \ref{sec:mitigate_yDMhost} introduces a technique to reduce the host signal.

\subsection{$y {\rm DM}$ value originating from a host}
\label{sec:yDM_host}

\begin{figure*}
 \begin{center}
   \includegraphics[width=17cm]{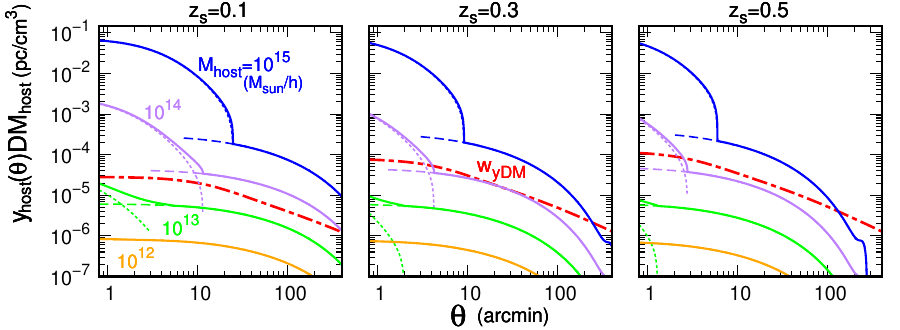}  %{fig_xi_host_paper.pdf} 
 \end{center}
\caption{A product of $y_{\rm host}$ and ${\rm DM}_{\rm host}$ obtained by \texttt{HMx}. The FRB is assumed to be located at the center of the host halo and the $y$ parameter is measured at an angular separation of $\theta$ from the center. The solid curves represent the results for host halo masses of $10^{15}$, $10^{14}$, $10^{13}$, and $10^{12} \, h^{-1} M_\odot$ from top to bottom. They are the sum of the 1-halo term (dotted curve) and the 2-halo term (dashed curve). 
\if0 The dotted curve ends at the virial radius of the halo.\fi The dot-dashed red curve is the cosmological cross-correlation $w_{y{\rm DM}}^{\rm (theo)}(\theta)$ in Subsection \ref{sec:HMx}. The amplitudes of all curves scale as $\propto (f_{\rm e}/0.9)^2$. 
%{Alt text: Line graphs demonstrating the substantial contribution of massive host halos to the cross-correlation, especially at small scales.}
}
\label{fig_xi_host}
\end{figure*}

%This subsection estimates the cross-correlation arising from a host, based on \texttt{HMx}.
%The bound gas density and temperature profiles in \texttt{HMx} are used to obtain ${\rm DM}_{\rm host}$ and $y_{\rm host}$. 
As a representative case, we consider an FRB located at the center of the host halo of mass $M_{\rm host}$.
Then, ${\rm DM}_{\rm host}$ is obtained by integrating the electron number density along a radial coordinate from the center to the virial radius $r_{\rm vir}$ (the ${\rm DM}$ outside $r_{\rm vir}$ is included in ${\rm DM}_{\rm cos}$).
The host contribution to the $y$ parameter, $y_{\rm host}$, comes from gas both inside and outside the halo (the 1- and 2-halo terms, respectively). 
When a CMB photon passes through the halo with an impact parameter (i.e., the closest distance to the center) of $\chi(z_{\rm s}) \theta$, the 1-halo term is derived from Eq.~(\ref{eq:y}):
\begin{align}
    y^{\rm 1h}_{\rm host}(\theta;M_{\rm host},z_{\rm s}) &= \frac{\sigma_{\rm T}}{m_{\rm e} c^2} \left( 1+z_{\rm s} \right)^2 \! \int_{-\infty}^{\infty} \!\!\! {\rm d}l  \nonumber \\
    &~\times p_{\rm e,host}(\sqrt{l^2+\chi(z_{\rm s})^2 \theta^2};M_{\rm host},z_{\rm s}),
\end{align}
where $p_{\rm e,host}(r;M_{\rm host},z_{\rm s})$ is the pressure profile of the halo (zero at $r > r_{\rm vir}$) and $l$ is the separation along the line-of-sight.
Because a halo forms at a local density peak in the large-scale structure, it is positively spatially correlated with the surrounding gas pressure even at $r > r_{\rm vir}$. 
The spatial cross-correlation between the halo number density contrast and the pressure perturbation is described in terms of its cross-power spectrum $P_{{\rm h},p_{\rm e}}(k;M_{\rm host},z_{\rm s})$. 
The 2-halo term is then obtained as\footnote{The three-dimensional cross-correlation is $\xi^{\rm 2h}_{{\rm h},p_{\rm e}}(r;M_{\rm host},z_{\rm s})= \int_0^\infty \! {\rm d}k \, k$ $P^{\rm 2h}_{{\rm h},p_{\rm e}}(k;M_{\rm host},z_{\rm s}) \, {\rm sin}(kr)/(2 \pi^2 r)$. Projecting this along the line-of-sight and multiplying by a factor of $\sigma_{\rm T} ( 1+z_{\rm s})^2/(m_{\rm e} c^2)$ yields Eq.~(\ref{eq:y_host_2h}).} (e.g., \cite{Li2011,Fang2012}) 
\begin{align}
    y^{\rm 2h}_{\rm host}(\theta;M_{\rm host},z_{\rm s}) &= \frac{\sigma_{\rm T}}{m_{\rm e} c^2} \left( 1+z_{\rm s} \right)^2 \! \int_0^{\infty} \!\!\frac{k \, {\rm d}k}{2 \pi} P^{\rm 2h}_{{\rm h},p_{\rm e}}(k;M_{\rm host},z_{\rm s}) \nonumber \\
    & ~~~~~~~~~~~~~~~~~~~~~~~~~~~~~~~~~~~\times J_0(k \chi(z_{\rm s}) \theta).
\label{eq:y_host_2h}
\end{align}
The 2-halo term of the cross-power spectrum is rewritten as $P^{\rm 2h}_{{\rm h},p_{\rm e}}(k;M_{\rm host},z_{\rm s}) = b_{\rm h}(M_{\rm host},z_{\rm s}) P^{\rm 2h}_{{\rm m},p_{\rm e}}(k;z_{\rm s})$, where $b_{\rm h}$ represents the linear halo bias obtained from \citet{Tinker2010}, and $P^{\rm 2h}_{{\rm m},p_{\rm e}}$ is the 2-halo term of the cross-power spectrum between the matter density contrast and the pressure perturbation in \texttt{HMx} (Appendix \ref{app:pk_norm}).

The product of $y_{\rm host}$ ($=y^{\rm 1h}_{\rm host}+y^{\rm 2h}_{\rm host}$) and ${\rm DM}_{\rm host}$ is
\begin{equation}
  (y {\rm DM})_{\rm host}(\theta;M_{\rm host},z_{\rm s}) \equiv y_{\rm host}(\theta;M_{\rm host},z_{\rm s}) \, {\rm DM}_{\rm host}(M_{\rm host},z_{\rm s}).
\label{eq:yDM_host}
\end{equation}
%The observable correlation is $\langle (y {\rm DM})_{\rm host} \rangle - \langle y_{\rm host} \rangle \langle {\rm DM}_{\rm host} \rangle$, resembling the cosmological correlation $w_{y{\rm DM}}^{\rm (theo)}$ in Eq.~(\ref{eq:w_yDM}).
%However, as the host halo population is unknown, the ensemble average $\langle \cdots \rangle$ cannot be obtained.
We calculate $(y {\rm DM})_{\rm host}$ in Eq.~(\ref{eq:yDM_host}) for several halo masses ($M_{\rm host}=10^{15}$, $10^{14}$, $10^{13}$, and $10^{12} \, h^{-1} M_\odot$) and redshifts ($z_{\rm s}=0.1$, $0.3$, and $0.5$) to examine its dependence on these parameters.
As shown in Fig.~\ref{fig_xi_host}, $(y {\rm DM})_{\rm host}$ is sensitive to the halo mass. 
Let us estimate the dependence of halo mass on $(y {\rm DM})_{\rm host}$ in the self-similar model. %while ignoring baryon feedback below. 
${\rm DM}_{\rm host}$ is proportional to the product of the gas density and the virial radius $r_{\rm vir}$. 
As the mean gas density within the halo is independent of $M_{\rm host}$ (determined by the virial over-density $\Delta_{\rm vir}$ times the cosmological background density at that epoch, where we implicitly assume that the FRB occurs at the same time as the halo formation) and $r_{\rm vir} \propto M_{\rm host}^{1/3}$, we have ${\rm DM}_{\rm host} \propto M_{\rm host}^{1/3}$. 
Similarly, $y_{\rm host}^{\rm 1h}$ is proportional to ${\rm DM}_{\rm host}$ multiplied by the halo virial temperature $T_{\rm vir}$. 
Using the virial theorem $k_{\rm B} T_{\rm vir} \propto M_{\rm host}/r_{\rm vir}$, we have $y_{\rm host}^{\rm 1h} \propto M_{\rm host}$; therefore, $(y^{\rm 1h} {\rm DM})_{\rm host} \propto M_{\rm host}^{4/3}$.
This estimate is consistent with the 1-halo term results in Fig.~\ref{fig_xi_host}.
In less massive halos of $M_{\rm host} \lesssim 10^{13} \, h^{-1} M_\odot$, because gas is effectively expelled through AGN feedback, both ${\rm DM}_{\rm host}$ and $y^{\rm 1h}_{\rm host}$ are further suppressed.
The halo mass dependence of $y^{\rm 2h}_{\rm host}$ arises from the halo bias. 
As the bias slightly increases with $M_{\rm host}$ (scaling approximately as $b_{\rm h} \propto M_{\rm host}^{1/3}$ for the mass and redshift ranges plotted in Fig.~\ref{fig_xi_host}), one obtains $(y^{\rm 2h} {\rm DM})_{\rm host} \propto M_{\rm host}^{2/3}$, which is roughly consistent with the results of the two-halo term in Fig.~\ref{fig_xi_host}. 
The 2-halo term exceeds the 1-halo term, especially at small halo masses ($M_{\rm host} \lesssim 10^{13}h^{-1}M_\odot$).

As shown in Fig.~\ref{fig_xi_host}, the host-halo contribution of massive halos ($M_{\rm host} \gtrsim 10^{14} \, h^{-1} M_\odot$) exceeds the cosmological cross-correlation, especially at small angles ($\theta \lesssim 10^\prime$).
Although these massive halos are rare, they will likely exert significant impact on the cross-correlation.
Whereas the cosmological cross-correlation $w_{y{\rm DM}}^{\rm (theo)}$ increases with $z_{\rm s}$, $(y {\rm DM})_{\rm host}$ is almost independent of $z_{\rm s}$.
In fact, $(y^{\rm 1h} {\rm DM})_{\rm host}(\theta=0;z_{\rm s})$ remains within a factor of $2$ in the range $z_{\rm s}=0$--$1$: at $z_{\rm s}=0.3$ and $\theta=0$, $(y^{\rm 1h} {\rm DM})_{\rm host}$ $\sim 0.1 \, {\rm pc}/{\rm cm}^3$ $(y^{\rm 1h}_{\rm host}/9 \times 10^{-5})$ $[{\rm DM}_{\rm host}/(1000 \, {\rm pc}/{\rm cm}^3)]$ for $M_{\rm host}=10^{15} \, h^{-1} M_\odot$ and $(y^{\rm 1h} {\rm DM})_{\rm host} \sim 4 \times  10^{-7} \, {\rm pc}/{\rm cm}^3$ $(y^{\rm 1h}_{\rm host}/2 \times 10^{-8})$ $[{\rm DM}_{\rm host}/(20 \, {\rm pc}/{\rm cm}^3)]$ for $M_{\rm host}=10^{12} \, h^{-1} M_\odot$.
%The angular size of $(y {\rm DM})_{\rm host}$ decreases as $z_{\rm s}$ increases (because it becomes more distant).

The above estimation assumes that the FRB resides at the halo center; the actual results depend on the FRB position within the halo.
Specifically, $(y{\rm DM})_{\rm host}$ will be larger (smaller) when the source is positioned behind (in front of) the center and/or is nearer (farther) the center in the transverse direction.  

\subsection{Cross-correlation arising from hosts}
\label{sec:yDM_host_ave}

\begin{figure*}
 \begin{center}
   \includegraphics[width=17cm]{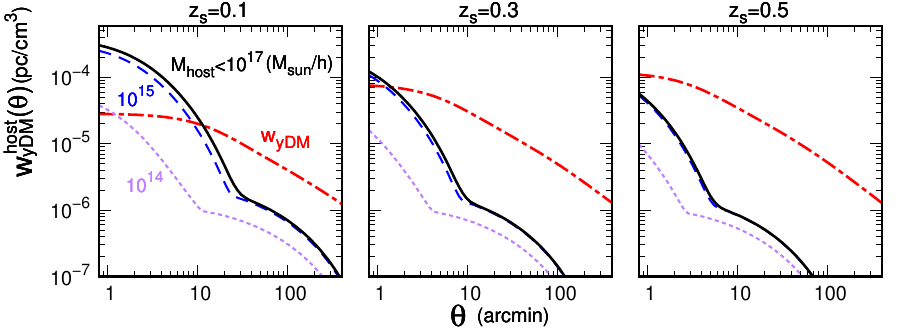}  %{fig_xi_host-ave_paper.pdf}  
 \end{center}
\caption{Similar to Fig.~\ref{fig_xi_host}, but the cross-correlations are calculated by ensemble-averaging host halo masses, assuming that the FRB rate traces stellar mass. The black curve shows the results for the default mass range $10^7$--$10^{17} h^{-1} M_\odot$. The dashed blue and dotted purple curves exclude halos with masses greater than $10^{15} h^{-1} M_\odot$ and $10^{14} h^{-1} M_\odot$, respectively. The dot-dashed red curve is the cosmological cross-correlation.  
%{Alt text: Similar to Fig.~\ref{fig_xi_host}, but showing ensemble-averaged results across host halo masses.}
}
\label{fig_xi_host-ave}
\end{figure*}

The host-induced cross-correlation $w_{y{\rm DM}}^{\rm host}(\theta;z_{\rm s})$ is obtained by taking an ensemble average over halo masses:
\begin{align}
    w_{y{\rm DM}}^{\rm host}(\theta;z_{\rm s}) \equiv& \langle (y {\rm DM})_{\rm host} \rangle (\theta;z_{\rm s})  \nonumber \\
    &- \langle y_{\rm host} \rangle (\theta;z_{\rm s}) \, \langle {\rm DM}_{\rm host} \rangle (z_{\rm s}).
\label{eq:w_yDM_host}
\end{align}
If we assume that the FRB rate traces stellar mass (e.g., \cite{Hashimoto2022,Sharma2024}), the ensemble average of a given function $Q$ can be written as
\begin{align}
    &\langle Q \rangle (\theta;z_{\rm s}) = \nonumber \\ &~~~\frac{\int {\rm d}M_{\rm host} M_{\rm host} \frac{{\rm d}n}{{\rm d}M}(M_{\rm host},z_{\rm s}) f_{*}(M_{\rm host}) Q(\theta;M_{\rm host},z_{\rm s})}{\int {\rm d}M_{\rm host} M_{\rm host} \frac{{\rm d}n}{{\rm d}M}(M_{\rm host},z_{\rm s}) f_{*}(M_{\rm host})},  
\label{eq:ave_by_mass}
\end{align}
where ${\rm d}n/{\rm d}M$ is the halo mass function~\citep{Tinker2008}, $f_{*}$ denotes the stellar mass fraction in \texttt{HMx} (Eq.~(27) of \citet{Mead2020}), and the integral is evaluated over the mass range $M_{\rm host} = 10^7$--$10^{17} h^{-1} M_\odot$. Figure \ref{fig_xi_host-ave} presents the corresponding results. This figure shows that, in general, the host contribution remains below the cosmological signal, except on smaller angular scales ($\theta \lesssim 10^\prime$) at lower redshifts ($z_{\rm s} < 0.1$). This smallness arises because (i) the mass function is exponentially suppressed at high masses ($M_{\rm host} \gtrsim$ a few $\times 10^{14} h^{-1} M_\odot$), and (ii) $f_{*}$ peaks at $M_{\rm host} = 10^{12.5} h^{-1} M_\odot$, where the host contribution is small (see Fig.~\ref{fig_xi_host}). When we exclude massive halos ($M_{\rm host} > 10^{14} h^{-1} M_\odot$) by changing the mass range of the integral in the numerator of Eq.~(\ref{eq:ave_by_mass}), the small-scale host contribution falls below the cosmological signal even at $z_{\rm s}=0.1$.

Let us consider how the host contribution, as discussed above, changes if the FRB rate traces the star formation rate (SFR) rather than stellar mass. In that case, because FRBs occur in less massive halos, the host contribution would be smaller. In addition, because FRBs occur at higher redshifts, their contribution would shift toward smaller angular scales.

\subsection{Mitigating the host contribution}
\label{sec:mitigate_yDMhost}

To minimize the host contribution, which contaminates the measurements of cosmological cross-correlation $w_{y{\rm DM}}$, we can 1) eliminate low-$z_{\rm s}$ sources in the cross-correlation analysis, 2) exclude small angular-scale signals, and/or 3) discard massive host-halo samples.
Regarding 1) and 2), we will discuss the dependencies of $z_{\rm s}$ and $\theta$ on the measured cross-correlation in Section \ref{sec:measurment}. 
Regarding 3), we searched for FRBs belonging to a cluster in the all-sky Planck catalog of SZ sources (PSZ2:~\cite{Planck2016_PSZ2,BH2024}), which includes 1,334 SZ clusters with masses\footnote{$M_{\rm 500c}$ is a spherical halo mass with an average density $500$ times higher than the cosmological critical density at that epoch.} $M_{\rm 500c} \gtrsim 10^{14} M_\odot$ and $z=0$--$1$. 
Three FRBs (20220914A, 20231206A, and 20231229A) satisfied the criteria of cluster-associated FRBs, namely, a redshift difference of $<0.03$ and a transverse separation of $< 3 \, h^{-1} {\rm Mpc}$ (corresponding to three times the virial radius of a $10^{14} M_\odot$ halo) between an FRB and a cluster. 
Two of them, 20220914A and 20231206A, reportedly belong to Abell 2310~\citep{Connor2023} and Abell 576~\citep{Amiri2025}, respectively. 
The host galaxy of 20231229A is UGC 1234~\citep{Amiri2025}, which belongs to Abell 262.
These three FRBs, hereafter referred to as ``cluster FRBs'', are excluded from our cross-correlation analysis but included in our ${\rm DM}$--$z$ analysis.

\section{Observational data}
\label{sec:obs_data}

This section summarizes our data on localized FRBs and $y$-maps. 

\subsection{Residual of the extragalactic ${\rm DM}$}
\label{sec:full-skymap_DeltaDMe}

\begin{table*}
 \tbl{Best-fit parameters derived from the ${\rm DM}$--$z$ relation with $131$ ($130$) FRBs for NE2001+YT20 (YMW16+YT20). 
The bold values indicate the maximum a posteriori (MAP) values of the posterior distribution (Eq.~\ref{eq:posterior}).
The values in parentheses represent the means $\pm$ $68\%$ credible intervals of the 1D marginalized posterior distributions.}
{\begin{tabular}{cccccc}
      \hline
       Model & $f_{\rm e}$ & ${\rm DM}_{\rm host,0}$ (${\rm pc/cm}^3$) & $\sigma_{\rm host,0}$ (${\rm pc/cm}^3$) & $\beta_{\rm host}$ & ${\rm DM}_{\rm MW}$ \\ 
      \hline
       ${\rm N}\beta$ & $\mathbf{0.972}$ $(0.918^{+0.082}_{-0.023})$ & $\mathbf{120.2}$ $(130.6^{+15.3}_{-24.0})$ & $\mathbf{98.9}$ $(114.9^{+16.8}_{-41.0})$ & $\mathbf{-0.199}$ $(-0.327^{+0.776}_{-1.070})$ & NE2001+YT20 \\
       ${\rm N}1$ & $\mathbf{1.000}$ $(0.972^{+0.027}_{-0.009})$ & $\mathbf{139.7}$ $(151.7^{+15.7}_{-21.0})$ & $\mathbf{122.8}$ $(147.2^{+24.1}_{-46.2})$ & $1$ & NE2001+YT20 \\
       ${\rm Y}\beta$ & $\mathbf{0.968}$ $(0.915^{+0.085}_{-0.024})$ & $\mathbf{124.1}$ $(136.6^{+14.9}_{-26.5})$ & $\mathbf{100.2}$ $(120.4^{+17.4}_{-43.1})$ & $\mathbf{-0.106}$ $(-0.177^{+0.833}_{-1.139})$ & YMW16+YT20 \\
       ${\rm Y}1$ & $\mathbf{1.000}$ $(0.970^{+0.029}_{-0.010})$ & $\mathbf{141.4}$ $(154.5^{+15.5}_{-21.2})$ & $\mathbf{122.0}$ $(147.4^{+23.5}_{-44.9})$ & $1$ & YMW16+YT20 \\
      \hline
    \end{tabular}}
    \label{table_bestfit_DM-z}
\begin{tabnote}
\footnotemark[]In the second and fourth rows, $\beta_{\rm host}$ is fixed at 1. 
%NE2001+YT20 is used for ${\rm DM}_{\rm MW}$ in the first two rows, while YMW16+YT20 is used in the last two rows.
\end{tabnote}
\end{table*}

\begin{figure}
 \begin{center}
   \includegraphics[width=8.5cm]{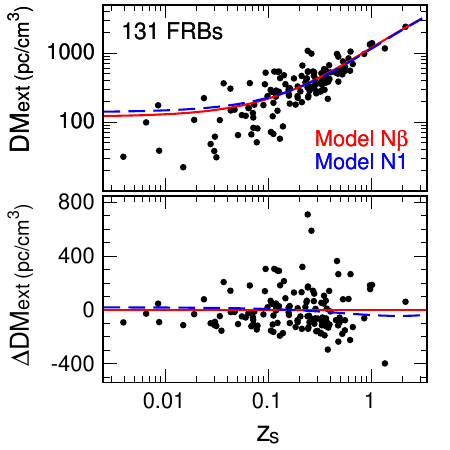}   %{fig_DM-z_paper.pdf} 
 \end{center}
\caption{The top panel shows the extragalactic ${\rm DM}$ ($\equiv {\rm DM}_{\rm obs}-{\rm DM}_{\rm MW}$) as a function of source redshift. 
The black-filled circles represent 131 FRBs.
The solid red and dashed blue curves represent the best-fitting theoretical models N$\beta$ and N1 in Table \ref{table_bestfit_DM-z}, respectively. 
The bottom panel displays the residual from the red curve. 
%{Alt text: In the upper panel, the amplitude of the theoretical curves is determined by the host dispersion measure at low redshifts and the ionized fraction at high redshifts, respectively.}
}
\label{fig_DM-z}
\end{figure}

The average ${\rm DM}_{\rm ext}$ at source redshift $z_{\rm s}$, $\overline{\rm DM}_{\rm ext}(z_{\rm s})$, is estimated from the ${\rm DM}$--$z$ relation with localized FRBs. 
The ${\rm DM}_{\rm ext}$ residual is defined by subtracting the average from ${\rm DM}_{\rm ext}$:
\begin{equation}
  \Delta {\rm  DM}_{\rm ext}(\bftheta;z_{\rm s}) \equiv {\rm  DM}_{\rm ext}(\bftheta;z_{\rm s}) - \overline{\rm  DM}_{\rm ext}(z_{\rm s}).
\label{eq:dDM_ext}
\end{equation}
Table \ref{table_list_FRBs1} of Appendix \ref{app:list_localized_FRB} lists the names, equatorial coordinates, ${\rm DM}_{\rm obs}$, and redshifts of the $133$ localized FRBs reported to date. %(the localized FRB data has recently been compiled into a numeric table in e.g., \cite{Connor2024} and \cite{Wang2025}).
Below, we provide sample estimates of $\overline {\rm DM}_{\rm ext}$.

We first explore the redshift dependence of ${\rm DM}_{\rm host}$.
If the host-galaxy property does not change over time in the rest frame, ${\rm DM}_{\rm host}$ decreases proportionally to $(1+z_{\rm s})^{-1}$ (e.g., \cite{Ioka2003,Zhou2014}). 
Previous theoretical studies examined the redshift evolution using hydrodynamic simulations, assuming that the FRB rate traces the stellar mass density or the SFR (e.g., \cite{Kovacs2024}).
These studies yielded varying results: an increase of ${\rm DM}_{\rm host}$ with redshift~\citep{Jaroszynski2020,Mo2023}, no significant evolution~\citep{Zhang2020,Kovacs2024}, or a slight decrease with redshift~\citep{Theis2024,Reischke2024}.
These differences come from variations in the models or assumptions used, including the FRB position in the host, the host-halo population, and baryon feedback.
%definition of halo radius when calculating it (i.e., the path length used for the integration of electron density)
%$\beta_{\rm host}=1$ (e.g., \cite{Zhou2014}). 
Therefore, we assume its redshift dependence as a simple power law of $1+z_{\rm s}$:
\begin{equation}
  {\rm DM}_{\rm host} = \frac{{\rm DM}_{\rm host,0}}{(1+z_{\rm s})^{\beta_{\rm host}}},
\label{eq:DMhost}
\end{equation}
where ${\rm DM}_{\rm host,0}$ is the host {\rm DM} at present and $\beta_{\rm host}$ is a free parameter. 

We assume that ${\rm DM}_{\rm cos}$, ${\rm DM}_{\rm host,0}$, and ${\rm DM}_{\rm MW}$ follow a log-normal distribution:
\begin{equation}
 P_{\rm LN}(x | \mu,\sigma) = \frac{1}{\sqrt{2 \pi} \sigma x} \exp \left[ - \frac{(\ln x -\mu)^2}{2 \sigma^2} \right],
\end{equation}
where the mean and standard deviation of $x$ are $e^{\mu+\sigma^2/2}$ and  $e^{\mu+\sigma^2/2} (e^{\sigma^2}-1)^{1/2}$, respectively. 
The mean and standard deviation of ${\rm DM}_{\rm cos}$ are obtained from Eqs.~(\ref{eq:DMcos}) and (\ref{eq:sigma_DMcos}), respectively, which are proportional to $f_{\rm e}$.
The standard deviation is computed using the electron power spectrum $P_{n_{\rm e}}$ in \texttt{HMx}.
The mean and standard deviation of ${\rm DM}_{\rm host,0}$ are given by $\overline{\rm DM}_{\rm host,0}$ and $\sigma_{\rm host,0}$, respectively. %with Eq.~(\ref{eq:DMhost}). 
The mean ${\rm DM}_{\rm MW}$ is NE2001+YT20 or YMW16+YT20, and the standard deviation is set to $0.5 \, {\rm DM}_{\rm MW}$\footnote{\citet{Price2021} estimated the accuracies of NE2001 and YMW16 using distance-known pulsars, which were excluded from the model calibrations of both models. They obtained a standard deviation of  $(0.5$--$0.6) \times {\rm DM}_{\rm NE2001/YMW16}$ (Fig.~6 in their paper) between the model prediction and the measured value.} (e.g., \cite{Hoffmann2025}).
FRBs with ${\rm DM}_{\rm ext}<0$ (two in NE2001+YT20 and three in YMW16+YT20) are excluded from the ${\rm DM}$--$z$ analysis and subsequent cross-correlation analysis. 
The ${\rm DM}$--$z$ relation can determine four parameters: $\bfp=(f_{\rm e}, \overline{\rm DM}_{\rm host,0}, \sigma_{\rm host,0}, \beta_{\rm host})$.
Let $z_{\rm s}^{(j)}$ and ${\rm DM}_{\rm obs}^{(j)}$ denote the redshift and observed ${\rm DM}$ of the $j$-th FRB, respectively. 
Here, the measurement error in ${\rm DM}_{\rm obs}^{(j)}$ is ignored because it is negligible (usually much smaller than $1 \, {\rm pc/cm}^3$).
The likelihood function of ${\rm DM}_{\rm obs}$ for all FRBs is (e.g., \cite{Macquart2020,Yang2022,Zhang2025})
\begin{equation}
    \mathcal{L}({\rm DM}_{\rm obs}|\bfp) = \prod_{j} %\in {\rm FRBs}}
    P({\rm DM}_{\rm obs}^{(j)}|\bfp),
\end{equation}
with
\begin{align}
     P({\rm DM}_{\rm obs}^{(j)}|\bfp) =& \int_0^{{\rm DM}_{\rm obs}^{(j)}} \!\!\!\!\!  {\rm d DM}_{\rm cos}  \int_0^{({\rm DM}_{\rm obs}^{(j)}-{\rm DM}_{\rm cos}) (1+z_{\rm s}^{(j)})^{\beta_{\rm host}}}  \hspace{-3.1cm} {\rm d DM}_{\rm host,0}  \nonumber \\
     &\times  P_{\rm LN}({\rm DM}_{\rm cos}|f_{\rm e}) \nonumber \\
     &\times  P_{\rm LN}({\rm DM}_{\rm host,0}| \overline{\rm DM}_{\rm host,0}, \sigma_{\rm host,0})   \nonumber \\  
     &\times P_{\rm LN} \left({\rm DM}_{\rm obs}^{(j)}-{\rm DM}_{\rm cos}-\frac{{\rm DM}_{\rm host,0}}{(1+z_{\rm s}^{(j)})^{\beta_{\rm host}}} \right).
\end{align}
The second, third, and fourth lines are the probability distributions of ${\rm DM}_{\rm cos}$, ${\rm DM}_{\rm host,0}$, and ${\rm DM}_{\rm MW}$, respectively.
The posterior probability distribution of $\bfp$ is defined by Bayesian inference:
\begin{equation}
  P(\bfp|{\rm DM}_{\rm obs}) \propto \mathcal{L}({\rm DM}_{\rm obs}|\bfp) \Pi(\bfp),
\label{eq:posterior}
\end{equation}
where $\Pi(\bfp)$ is the prior distribution.
We adopt a flat prior within the ranges $f_{\rm e}=[0,1]$, ${\rm DM}_{\rm host,0} = \sigma_{\rm host,0}=[0,400] \,{\rm pc/cm}^3$, and $\beta_{\rm host}=[-4,4]$, and perform Markov Chain Monte Carlo (MCMC) sampling using {\texttt{emcee}}~\citep{EMCEE2013}.
Table \ref{table_bestfit_DM-z} presents the best-fit values derived from the posterior distribution (\ref{eq:posterior}) using {\texttt{GetDist}}~\citep{Lewis2019}.
The maximum a posteriori (MAP) values will be used in the following cross-correlation analysis.
The values in parentheses represent the means and $68 \%$ credible intervals of the 1D marginalized posterior distributions.
The slight differences between the MAP and mean values are primarily attributed to projection of the posterior.
In the first and third rows of Table \ref{table_bestfit_DM-z}, $\beta_{\rm host}$ is nearly zero (despite the large credible interval), suggesting that ${\rm DM}_{\rm host}$ does not significantly evolve with redshift\footnote{Very recently, while we were preparing this paper, \citet{Acharya2025} similarly analyzed 65 localized FRBs and found a result ($\beta_{\rm host} \simeq 0$--$1$) consistent with ours.}. 
For negative $\beta_{\rm host}$, both ${\rm DM}_{\rm cos}$ and ${\rm DM}_{\rm host}$ increase with $z_{\rm s}$ but with different redshift dependencies, especially at low $z_{\rm s}$; specifically, ${\rm DM}_{\rm cos} \propto z_{\rm s}$ while ${\rm DM}_{\rm host} \propto (1+z_{\rm s})^{-\beta_{\rm host}}$. Therefore, each fitting parameter ($f_{\rm e}$, ${\rm DM}_{\rm host,0}$, and $\beta_{\rm host}$) can be determined almost independently.
$f_{\rm e}$ is somewhat smaller in the first and third rows than in the second and fourth rows, indicating that a larger ${\rm DM}_{\rm host}$ compensates for a smaller ${\rm DM}_{\rm cos}$ at higher redshifts.
Additional information, such as the scattering time, would tighten the constraints on the host property (e.g., \cite{Cordes2022,Yang2025}).

Figure \ref{fig_DM-z} plots ${\rm DM}_{\rm ext}$ as a function of redshift (the same plot in a linear-linear scale is Fig.~\ref{fig_DM-z_linear} in Appendix \ref{sec:DM-z_linear}).
The near-overlap of the solid red and dashed blue curves, calculated using the MAP values in the top two rows of Table \ref{table_bestfit_DM-z}, indicate that the current samples can be fitted by either model, although the N$\beta$ model with an extra free parameter $\beta_{\rm host}$ more accurately traces the redshift evolution.
The bottom panel presents the residual from the red curve, which will be correlated with the $y$ parameter.
The two largest $\Delta {\rm DM}_{\rm ext}$ are $711$ and $590 \, {\rm pc}/{\rm cm}^3$ from 20190520B (at $z_{\rm s}=0.241$) and 20220831A (at $z_{\rm s}=0.262$), respectively, while the two smallest $\Delta {\rm DM}_{\rm ext}$ are $-398$ and $-295 \, {\rm pc}/{\rm cm}^3$ from 20230521B (at $z_{\rm s}=1.354$) and 20190611B (at $z_{\rm s}=0.3778$), respectively.

\subsection{Compton $y$-maps}
\label{ssec:ymap}

\begin{figure*}[t]
  %\begin{minipage}{0.5\linewidth}
        \begin{center}
        \includegraphics[width=12cm]{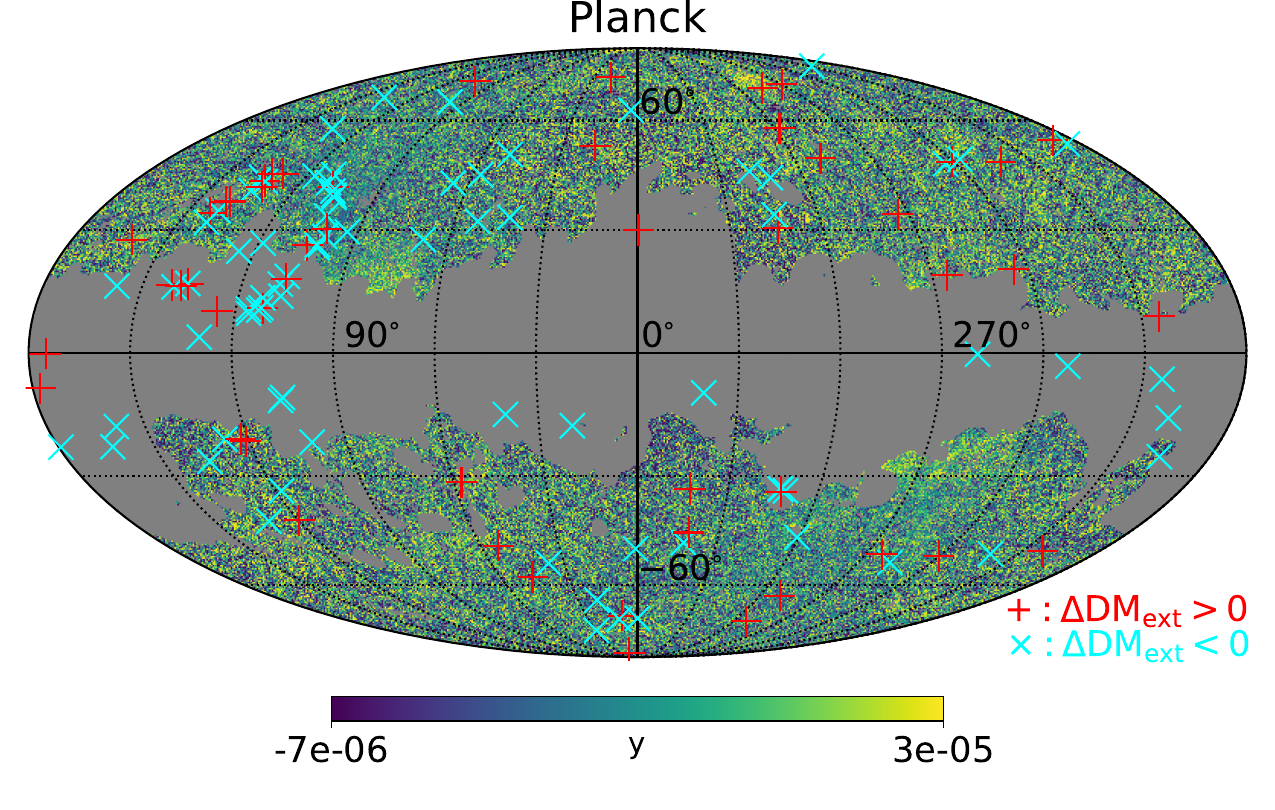}  %{fig_planck_map.pdf}
        %\end{center}
  %\end{minipage}
  %\begin{minipage}{0.5\linewidth}
        %\begin{center}
        \includegraphics[width=12cm]{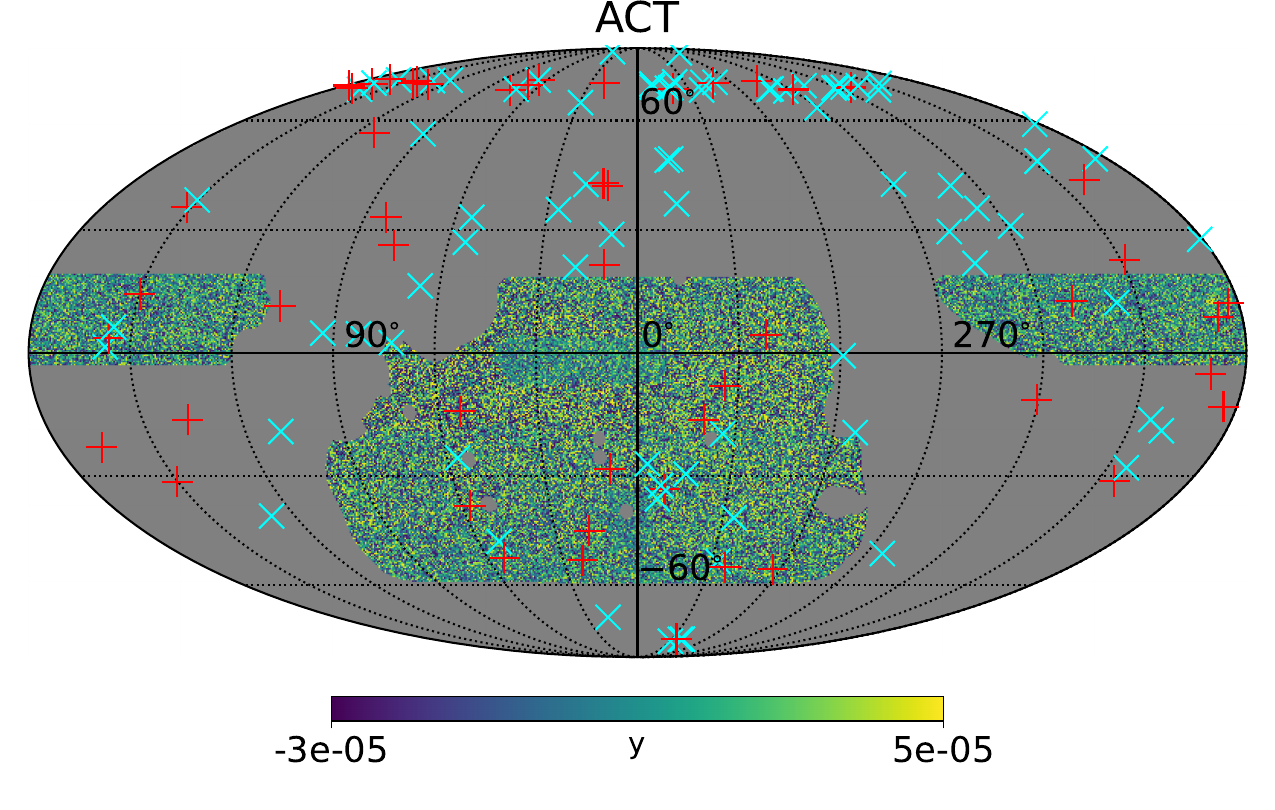}   %{fig_ACT_map.pdf}
        \end{center}
  %\end{minipage}
\caption{Compton $y$-maps from Planck PR2 MILCA (top panel) and ACT DR6 (bottom panel). The red plus and cyan cross symbols indicate FRBs with positive and negative $\Delta {\rm DM}_{\rm ext}$, respectively, in the N$\beta$ model. The top and bottom panels are displayed in galactic and equatorial coordinates, respectively.
%{Alt text: $y$-maps overlaid with source positions.}
}
\label{fig_maps}
\end{figure*}

This subsection briefly overviews the $y$-maps from the Planck Public Data Release 2 (PR2) and ACT Data Release 6 (DR6).

\subsubsection{Planck}

The Planck PR2 data include full-sky maps from nine frequency channels (30 to 857 GHz) collected between August 2009 and October 2013~\citep{Planck2016_overview}. 
The $y$-map was extracted from these maps using the characteristic frequency dependence of the tSZ effect~\citep{Planck2016_ymap}. 
Planck PR2 offers two $y$-maps\footnote{\url{https://irsa.ipac.caltech.edu/data/Planck/release_2/all-sky-maps/ysz_index.html}\label{ftn_P}} (including the standard deviation of the noise) obtained through different methods: Needlet Independent Linear Combination (NILC) and Modified Internal Linear Combination Algorithm (MILCA).
%Both methods are based on the Internal Linear Combination (ILC) technique.
Planck PR2 also provides foreground masks of the Galactic plane and bright point sources.
Combining the Galactic mask, which excludes $40 \%$ of the sky around the Galactic disk, with the point-source mask, the $y$-map covers $\sim 51\%$ of the sky.
All of these maps are provided in the Healpix scheme with $N_{\rm side}=2048$~\citep{Gorski2005}, corresponding to a pixel size of $\sim 2^\prime$.
The beam size of the $y$-map is assumed to follow a Gaussian distribution with a full width at half maximum (FWHM) of $10^\prime$.

Figure \ref{fig_maps} (top panel) presents the MILCA $y$-map overlaying the FRB positions.
The survey region includes $71$ FRBs with an average redshift of $0.27$ (excluding the three cluster FRBs as discussed in Subsection \ref{sec:mitigate_yDMhost}).

\subsubsection{ACT}

The Atacama Cosmology Telescope is located in the Atacama Desert of Chile.
The DR6 data include three frequency maps at 93, 148, and 225 GHz collected from 2017 to 2022.
These maps and the Planck maps at eight frequencies (30 to 545 GHz) were combined to construct the $y$-map using the NILC pipeline~\citep{Coulton2024}.
The $y$-map and mask, provided in equatorial coordinates by the ACT team\footnote{\url{https://lambda.gsfc.nasa.gov/product/act/actadv_dr6_compton_maps_info.html}}, were transformed into the Healpix format with $N_{\rm side}=8192$ (pixel size $\sim 0^\prime.5$). 
After removing the masked region, the $y$-map covers $\sim 34 \%$ of the sky.
The beam size is $1^\prime.6$ FWHM, significantly improved from that of Planck.

The $y$-map is plotted in the bottom panel of Fig.~\ref{fig_maps}. 
The survey region includes $31$ FRBs with an average redshift of $0.32$.
Twenty-five of these FRBs also reside in the Planck survey area (Table \ref{table_list_FRBs1}).
No cluster FRBs exist in the survey region.
The numerous FRBs clustered at ${\rm DEC} \simeq 70^\circ$ are attributed to DSA-110.

\section{Cross-correlation measurements}
\label{sec:measurment}

This section presents an estimator for the angular cross-correlation function (Subsection \ref{sec:w_yDM_estimator}) and the measurement results (Subsections \ref{sec:w_yDM_results}--\ref{sec:constraint_Te}). 

\subsection{A cross-correlation estimator}
\label{sec:w_yDM_estimator}

Let $\Delta {\rm DM}_{\rm ext}^{(j)}$ represent $\Delta {\rm DM}_{\rm ext}$ of the $j$-th FRB in the direction $\bftheta_{\rm FRB}^{(j)}$ and let $\bftheta_y$ be an angular position on the $y$-map (i.e., pixel coordinates in Healpix).
Using the FRBs within the survey region of the $y$-map, we calculate the cross-correlation by summing all pairs of $\Delta {\rm DM}_{\rm ext}^{(j)}$ and $y$ based on their angular separation $|\bftheta_{\rm FRB}^{(j)}-\bftheta_y|$.
An estimator of the correlation function is 
\begin{align}
      \widehat{w}_{y{\rm DM}}(\theta) &= \frac{\sum_{j,\bftheta_y} w_j(\bftheta_y) \Delta {\rm DM}_{\rm ext}^{(j)} \, \delta y(\bftheta_y)}{\sum_{j,\bftheta_y} w_j(\bftheta_y)}
     \nonumber \\ &~~ -\left. \frac{\sum_{j,\bftheta_y} w_j(\bftheta_y) \Delta {\rm DM}_{\rm ext}^{(j)} \, \delta y(\bftheta_y)}{\sum_{j,\bftheta_y} w_j(\bftheta_y)} \right|_{\rm random},
\label{eq:w_yDM_estimator}
\end{align}
where $\delta y(\bftheta_y)=y(\bftheta_y)-\bar{y}$ and $\bar{y}$ is the average $y$ in the survey region.
The summation is calculated when $\theta-\Delta \theta/2 \leq |\bftheta_{\rm FRB}^{(j)}-\bftheta_y| < \theta+\Delta \theta/2$ with a bin-width of $\Delta \log_{10} \theta=0.25$.
Here, $|\bftheta_{\rm FRB}^{(j)}-\bftheta_y|$ ranges from $1^\prime$ to $1000^\prime$.
The denominator of the first term is obtained in the same way as the numerator, but setting $\Delta {\rm DM}_{\rm ext}^{(j)}=\delta y=1$.
The estimator provides the average excess of $\Delta {\rm DM}_{\rm ext} \delta y$ within an annulus of radius $\theta$ and width $\Delta \theta$ around the FRBs.
For the weight function $w_j$, the inverse variance weight\footnote{This weight is optimal for galaxy-galaxy lensing, offering the highest signal-to-noise ratio in the cross-correlation between foreground galaxies and background weak-lensing shear when shot noise dominates the covariance (\cite{ShirasakiTakada2018} and references therein).} is employed:
\begin{equation}
   w_j(\bftheta_y) = \left[ \left( \frac{{\rm DM}_{\rm MW}^{(j)}}{2} \right)^2 + \left( \frac{\sigma_{{\rm host},0}}{(1+z_{\rm s}^{(j)})^{\beta_{\rm host}}} \right)^2 \right]^{-1} \!\!\!\!\! \sigma_y^{-2}(\bftheta_y),
\label{eq:weight_w_yDM}
\end{equation}
when both $\bftheta_{\rm FRB}^{(j)}$ and $\bftheta_y$ are in the survey region of the $y$-map; otherwise, $w_j(\bftheta_y)=0$. 
The first and second terms of Eq.~(\ref{eq:weight_w_yDM}) represent the variances of ${\rm DM}_{\rm MW}$ and ${\rm DM}_{\rm host}$, respectively; the second term down-weights the lower-redshift FRBs, for which the ${\rm DM}_{\rm host}$ variance exceeds the ${\rm DM}_{\rm cos}$ variance.
In the last term, $\sigma_y^2$ is the noise variance of $y$, assigned using public data for Planck and set to $\sigma_y=1$ for ACT. 

The second line in Eq.~(\ref{eq:w_yDM_estimator}) is the same as the first line but represents the correlation between randomly positioned FRBs and the $y$-map.
Here, the FRB's angular positions are randomly relocated within the survey region without changing their $\Delta {\rm DM}_{\rm ext}$.
%We randomly change the FRB's galactic longitudes (i.e., keeping their galactic latitudes) so that the galactic foreground level stays almost the same, following \citet{Tanimura2019}.
The second line is computed as the average of $3000$ iterations of this procedure.
The result is very small, typically less than a few percent of the first line. 
If $\Delta {\rm DM}_{\rm ext}$ and $\delta y$ do not correlate, the second line should ensure that the estimator value becomes zero.

The covariance of the cross-correlation is estimated through jackknife resampling (e.g., \cite{Norberg2009}). 
For $N_{\rm FRB}$ sources in the survey area, one source is removed at each time and the estimator is calculated with the remaining $N_{\rm FRB}-1$ sources. 
This process is repeated for all sources, yielding $N_{\rm FRB}$ correlations from which the covariance is determined. 
The covariance estimate is cross-checked using the bootstrap method. 
The estimator is obtained by randomly selecting $N_{\rm FRB}$ sources (allowing duplicates) in the survey region. 
The covariance is calculated after obtaining $3000$ correlations by repeating the above process.

\subsection{Measurement results}
\label{sec:w_yDM_results}

\begin{figure*}[t]
        \begin{center}
        \includegraphics[width=16cm]{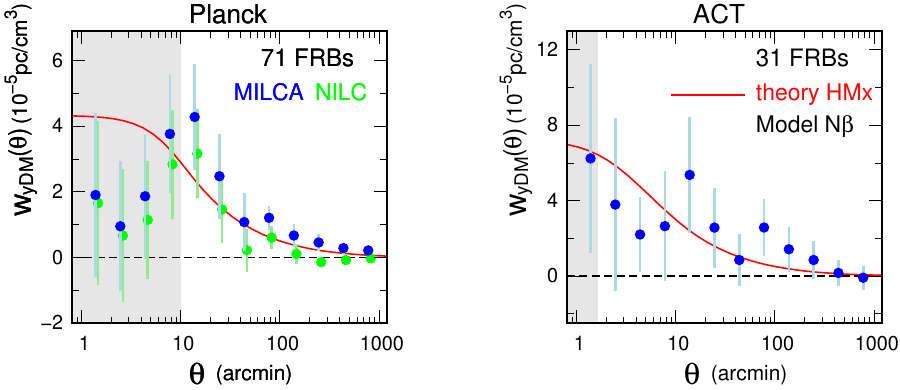}   %{fig_xi_Planck-ACT_paper.pdf}
        \end{center}
\caption{Cross-correlation measurements for Planck (left panel; $71$ FRBs) and ACT (right panel; $31$ FRBs). 
The filled circles with error bars represent the measurements with their standard deviations and the red curve depicts the theoretical prediction of \texttt{HMx}. 
Notably, the red curve does {\textit{not}} fit the cross-correlation measurements; the amplitude ($w_{y{\rm DM}}^{\rm (theo)} \propto f_{\rm e}^2$) is determined by the DM--$z$ relation (Fig.~\ref{fig_DM-z}).
Both panels use the N$\beta$ model in Table \ref{table_bestfit_DM-z}.
The shaded areas indicate the angular scales within the beam size of each detector, %($10^\prime$ and $1^\prime.6$ for Planck and ACT, respectively), 
which are excluded from the analysis.
The results in the left panel are slightly offset along the $x$-axis for visual clarity.
%{Alt text: The theoretical curve fairly agrees with the data points.}
}
\label{fig_xi_obs}
\end{figure*}

\begin{table}
  \tbl{Constraints on the amplitude $\mathcal{A}$ of the cross-correlation (means with $68 \%$ credible intervals). Here, $\mathcal{A} = 1$ corresponds to the \texttt{HMx} theoretical prediction.}  % for each model.}
%The fitting range is $\theta=10^\prime$--$316^\prime$ for Planck and $\theta=1^\prime.78$--$316^\prime$ for ACT.}
{%
 \begin{tabular}{ccc}
      \hline
      Model &  Planck MILCA & ACT \\
      \hline
       N$\beta$ & $2.01 \pm 0.50$ & $1.23 \pm 0.82$ \\
       N1       & $1.80 \pm 0.45$ & $0.87 \pm 0.78$ \\
       Y$\beta$ & $1.89 \pm 0.50$ & $1.18 \pm 0.83$ \\
       Y1       & $1.72 \pm 0.45$ & $0.79 \pm 0.81$ \\
       \hline
    \end{tabular}}
    \label{table:amp}
\begin{tabnote}
\end{tabnote}
\end{table}

\begin{table}
  \tbl{Same as Table \ref{table:amp} (the N$\beta$ model), but showing the constraints at different minimum source redshifts $z_{\rm s,min}$. $N_{\rm FRB}$ indicates the number of sources.}
%The fitting range is $\theta=10^\prime$--$316^\prime$ for Planck and $\theta=1^\prime.78$--$316^\prime$ for ACT.}
{%
 \begin{tabular}{cccccc}
      \hline
      & \multicolumn{2}{c}{Planck MILCA} & & \multicolumn{2}{c}{ACT} \\
      \cline{2-3} \cline{5-6}
      %\hline
      $z_{\rm s,min}$ & $\mathcal{A}$ & $N_{\rm FRB}$ & & $\mathcal{A}$ & $N_{\rm FRB}$  \\ 
      \hline
       0   & $2.01 \pm 0.50$ & 71 & & $1.23 \pm 0.82$ & 31 \\
       0.1 & $2.09 \pm 0.55$ & 56 & & $1.47 \pm 0.87$ & 26 \\
       0.2 & $1.90 \pm 0.61$ & 34 & & $1.23 \pm 1.57$ & 17 \\
       0.3 & $2.12 \pm 0.67$ & 23 & & ---- & 13 \\
%       0.4 & $1.47 \pm 0.61$ & 13 & & ----$\dag$ & 6 \\
       \hline
    \end{tabular}}
    \label{table:amp_zmin}
\begin{tabnote}
For ACT at $z_{\rm s,min} = 0.3$, the inverse covariance and the resulting constraint cannot be determined because the \citet{Hartlap2007} factor becomes infinite due to the limited number of realizations (from $N_{\rm FRB} = 13$).
\end{tabnote}
\end{table}

Figure \ref{fig_xi_obs} presents the cross-correlation measurements of Planck (left panel) and ACT (right panel) using the estimator in Eq.~(\ref{eq:w_yDM_estimator}). 
The error bars represent the standard deviations calculated using the jackknife method (Subsection \ref{sec:w_yDM_estimator}).
The jackknife and bootstrap estimates agree within $9 \%$.
The errors are strongly correlated, especially between nearby angular separations (see Fig.~\ref{fig_xi-cov} in Appendix \ref{app:cov}).
The red curve is the theoretical cosmological cross-correlation (Subsection \ref{sec:HMx}), including Gaussian smoothing with $10^\prime$ and $1^\prime.6$ FWHM for Planck and ACT, respectively.
The theoretical curve includes the same weight (by setting $\sigma_y=1$) as the measurements in Eq.~(\ref{eq:weight_w_yDM}); $w_{y {\rm DM}}^{\rm (theo)}(\theta) = \sum_j w_j w_{y {\rm DM}}^{\rm (theo)}(\theta;z_{\rm s}^{(j)})/\sum_j w_j$, where the summation is calculated over all FRBs in the correlation measurement.
The theoretical correlation depends on $f_{\rm e}$, $\sigma_{\rm host,0}$, and $\beta_{\rm host}$ (the last two parameters are included in the weight $w_j$).
These parameters were determined using the ${\rm DM}$--$z$ relation in Table \ref{table_bestfit_DM-z}.
The red curve apparently agrees with the measurements, even though it was not fitted to the correlation data.
The apparently larger correlation amplitude for Planck MILCA than for NILC at $\theta \gtrsim 10^\prime$ is attributable to large-scale noise at multipoles $\ell \lesssim 100$ in the NILC map (Fig.~5 of \cite{Planck2016_ymap}; also \cite{Vikram2017}).
Hereafter, we show only the results of the MILCA map for Planck.
Figure \ref{fig_xi_obs} uses the N$\beta$ model in Table \ref{table_bestfit_DM-z}.
The choice of model slightly influences the measurement results and the theoretical predictions, as discussed in the next paragraph.

To estimate the agreement between the theory and measurements, we substitute the amplitude of the theoretical cross-correlation as 
\begin{equation}
    w_{y{\rm DM}}^{\rm (theo)}(\theta)  \rightarrow \mathcal{A} \, w_{y{\rm DM}}^{\rm (theo)}(\theta)
\end{equation}
and analyze the likelihood of $\mathcal{A}$ assuming a Gaussian likelihood function of the cross-correlation:
\begin{align}
        \ln \mathcal{L}(\widehat{w}_{y{\rm DM}}|\mathcal{A}) =& -\frac{1}{2} \sum_{\theta,\theta^\prime} {\rm Cov}^{-1} (\theta,\theta^\prime) \left( \widehat{w}_{y{\rm DM}}(\theta) - \mathcal{A} \, w_{y{\rm DM}}^{\rm (theo)}(\theta) \right)  \nonumber \\
        &~~~~~~\times \left( \widehat{w}_{y{\rm DM}}(\theta^\prime) - \mathcal{A} \, w_{y{\rm DM}}^{\rm (theo)}(\theta^\prime) \right),
\label{eq:likeli_amp}
\end{align}
where ${\rm Cov}$ is the covariance matrix of the cross-correlation, obtained from the jackknife. % (Subsection \ref{sec:w_yDM_estimator}).
The inverse matrix of ${\rm Cov}$ incorporates the \citet{Hartlap2007} correction factor.
Similarly to the measurements (Eq.~\ref{eq:w_yDM_estimator}), the theoretical correlation function is also binned into $\theta$ bins. 
The summation in Eq.~(\ref{eq:likeli_amp}) is calculated over the range $\theta=10^\prime$--$1000^\prime$ for Planck and $\theta=1^\prime.78$--$1000^\prime$ for ACT (excluding the small angular scale of each detector’s beam size).
The mean and standard deviation of $\mathcal{A}$ are then given by (e.g., \cite{BICEP2016,Namikawa2019})
\begin{align}
    \bar{\mathcal{A}} &= \frac{\sum_{\theta,\theta^\prime} {\rm Cov}^{-1} (\theta,\theta^\prime) \, \widehat{w}_{y{\rm DM}}(\theta) \, w_{y{\rm DM}}^{\rm (theo)}(\theta^\prime)}{\sum_{\theta,\theta^\prime} {\rm Cov}^{-1} (\theta,\theta^\prime) \, w_{y{\rm DM}}^{\rm (theo)}(\theta) \, w_{y{\rm DM}}^{\rm (theo)}(\theta^\prime)}, \nonumber \\
    \sigma_\mathcal{A} &= \left[ \sum_{\theta,\theta^\prime} {\rm Cov}^{-1} (\theta,\theta^\prime) \, w_{y{\rm DM}}^{\rm (theo)}(\theta) \, w_{y{\rm DM}}^{\rm (theo)}(\theta^\prime) \right]^{-1/2}.
\end{align}
Table \ref{table:amp} lists $\bar{\mathcal{A}}$ and $\sigma_\mathcal{A}$ for the four models in Table \ref{table_bestfit_DM-z}.
The confidence level of nonzero detection is $(3.8$--$4.0) \, \sigma$ for Planck and $(1.0$--$1.5) \, \sigma$ for ACT, depending on the model.
The N$\beta$ and Y$\beta$ models predict a somewhat larger amplitude than N1 and Y1 but all results are consistent within the $1 \sigma$ confidence level.
The N$\beta$ and Y$\beta$ models, as well as N1 and Y1, each pair predict nearly the same result, indicating that the choice of ${\rm DM}_{\rm MW}$ model (NE2001 or YMW16) does not influence the cross-correlation measurements.
Hereafter, the N$\beta$ model will serve as the default unless stated otherwise.

The systematically smaller amplitude $\mathcal{A}$ for ACT than for Planck arises from fitting different angular ranges. 
When both datasets are fitted over the same angular range $\theta=10^\prime$--$1000^\prime$, the amplitude of ACT becomes $\mathcal{A} = 1.92 \pm 0.92$, comparable to that of Planck $\mathcal{A} = 2.01 \pm 0.50$.  
The small measurement signals at $\theta \leq 10^\prime$ lower the value of $\mathcal{A}$.
Another reason for the ACT's smaller $\mathcal{A}$ is that it yields a slightly lower correlation near $\theta = 10^\prime$ than Planck, even when using the same $25$ FRBs that lie in both detectors’ survey regions under the same measurement conditions (discussion in Appendix \ref{app:consistency}).

The theoretical prediction based on the flat-sky approximation becomes less accurate at larger angles (close to $\theta \approx 1 \, {\rm rad}$). 
However, after excluding the large angular signals at $\theta > 100^\prime$, the constraint remains almost unchanged---$\mathcal{A} = 1.89 \pm 0.59$ for Planck and $\mathcal{A} = 1.12 \pm 0.76$ for ACT---because the positive amplitude $\mathcal{A}$ is mainly contributed by the signal in $\theta=10^\prime$--$100^\prime$.

We also exclude the lower-redshift FRBs from the $\mathcal{A}$ estimation, as the hosts of these FRBs may contribute to the correlation signal (Section \ref{sec:host_galaxy_corr}). 
Table \ref{table:amp_zmin} lists the constraints on $\mathcal{A}$ at several minimum source redshifts $z_{\rm s,min}$. 
At higher $z_{\rm s,min}$, the constraint is weakened by the limited number of sources.
The constraint is insensitive to minimum redshifts $z_{\rm s,min} \leq 0.3$ because 1) higher-redshift sources yield stronger correlation signals and 2) the weight (Eq.~\ref{eq:weight_w_yDM}) in the estimator reduces the contribution from lower-redshift sources.
If the correlation includes substantial host contribution from nearby sources, $\mathcal{A}$ decreases with $z_{\rm s,min}$, but such a trend is absent in Table \ref{table:amp_zmin}. 
Therefore, we believe the host contribution is insignificant in the current measurements.

We note that as ${\rm DM}_{\rm cos} \propto f_{\rm e}$ in the ${\rm DM}$--$z$ relation and $w_{y{\rm DM}} \propto f_{\rm e}^2$ in the correlation, combining these measurements will obtain a more precise determination of $f_{\rm e}$ when the cross-correlation is measured more accurately.

\subsection{FRBs contributing to the positive cross-correlation}
\label{sec:w_yDM_topFRBs}

\begin{table}
\tbl{The top five FRBs generating the largest cross-correlations for Planck over different angular ranges: $\theta = 1^\prime$--$10^\prime$ (upper panel) and $\theta = 10^\prime$--$100^\prime$ (lower panel).}
%The Planck MILCA and model N$\beta$ are used.
%The second column is the cross-correlation for the Planck MILCA, using model N$\beta$.
{%
 \begin{tabular}{lccl}
       $\theta = 1^\prime$--$10^\prime$ & & & \\
       \hline
       ~~~~~FRB & $\widehat{w}_{y{\rm DM}}$ $({\rm pc}/{\rm cm}^3)$ & $\Delta {\rm DM}_{\rm ext}$ $({\rm pc}/{\rm cm}^3)$ & ~~~$z_{\rm s}$\\ 
      \hline
       20220914A$^\dag$ & $4.70 \times 10^{-4}$ & ~~~$306.2$ & $0.1139$ \\
       20220224C & $4.66 \times 10^{-4}$ & ~~~$266.8$ & $0.6271$ \\
       20220529A & $3.66 \times 10^{-4}$ & $-126.4$ & $0.1839$ \\
       20240114A & $3.24 \times 10^{-4}$ & ~~~$183.3$ & $0.13$ \\
       20240310A & $3.02 \times 10^{-4}$ & ~~~$291.6$ & $0.127$ \\
       %20180924B & $2.57 \times 10^{-4}$ & $-163.5$ & 0.3214 \\
      \hline
      \\
      $\theta = 10^\prime$--$100^\prime$ & & & \\    
      \hline
       ~~~~~FRB & $\widehat{w}_{y{\rm DM}}$ $({\rm pc}/{\rm cm}^3)$ & $\Delta {\rm DM}_{\rm ext}$ $({\rm pc}/{\rm cm}^3)$ & ~~~$z_{\rm s}$ \\ 
      \hline
       20231206A$^\dag$ & $2.16 \times 10^{-4}$ & ~~~$182.1$ & $0.0659$ \\
       20240310A & $1.32 \times 10^{-4}$ & ~~~$291.6$ & $0.127$ \\
       20240114A & $1.02 \times 10^{-4}$ & ~~~$183.3$ & $0.13$ \\
       20231025B & $7.70 \times 10^{-5}$ & $-171.2$ & $0.3238$ \\
       20220224C & $7.11 \times 10^{-5}$ & ~~~$266.8$ & $0.6271$ \\
       %20231230A & $6.61 \times 10^{-5}$ & $-115.7$ & 0.0298 \\
      \hline
    \end{tabular}}
    \label{table:Top5}
\begin{tabnote}
\footnotemark[$\dag$] the cluster FRBs
\end{tabnote}
\end{table}

\begin{figure}
 \begin{center}
   \hspace*{-1cm}
   \includegraphics[width=10cm]{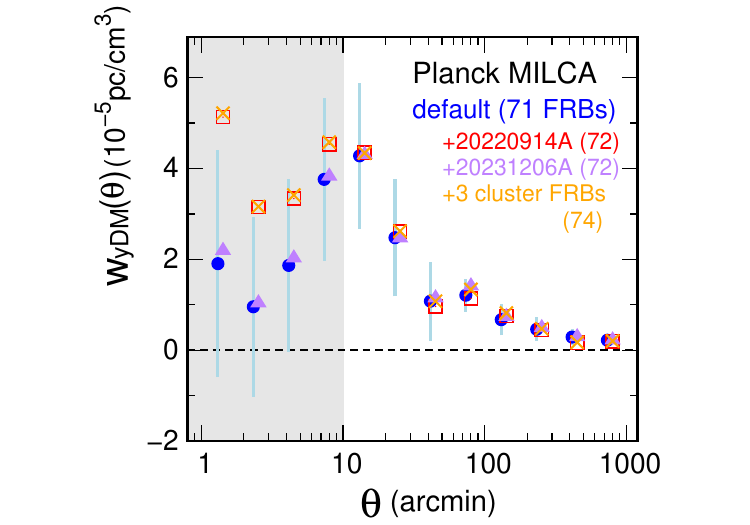}   %{fig_xi_top-FRBs_paper.pdf} 
 \end{center}
\caption{Contributions of the cluster FRBs to the cross-correlation measurement. 
The blue symbols with error bars represent our default. 
The red squares and purple triangles are the results of adding 20220914A or 20231206A to the default, respectively. 
The orange crosses are the result of adding the three cluster FRBs.  
The values in parentheses indicate the number of FRBs used in the analysis. 
The blue symbols are slightly offset along the $x$-axis for visual clarity.
%{Alt text: The host-cluster contribution is only apparent at small scales.}
}
\label{fig_xi_top-FRBs}
\end{figure}

\begin{table}
\tbl{The top five FRBs contributing to $\mathcal{A}$. The second column shows $\mathcal{A}$ when each source is excluded from the $74$-source set (the default $71$ and the three cluster FRBs, for which $\mathcal{A}=2.15 \pm 0.52$). A smaller $\mathcal{A}$ indicates a larger contribution.}
{%
 \begin{tabular}{lccl}
       \hline
       ~~~~~FRB & $\mathcal{A}$ & $\Delta {\rm DM}_{\rm ext}$ $({\rm pc}/{\rm cm}^3)$ & ~~~$z_{\rm s}$\\ 
      \hline
       20231025B & $0.90 \pm 0.33$ & $-171.2$ & $0.3238$ \\
       20220418A & $1.88 \pm 0.51$ & $-223.4$ & $0.622$  \\
       20240310A & $1.89 \pm 0.49$ & ~~~$291.6$ & $0.127$ \\
       20231020B & $2.02 \pm 0.52$ & ~~~$266.7$ & $0.4775$ \\
       20180924B & $2.02 \pm 0.52$ & $-174.4$ & $0.3214$ \\ 
      \hline
    \end{tabular}}
    \label{table:Top5A}
\begin{tabnote}
%\footnotemark[$\dag$] the cluster FRBs
\end{tabnote}
\end{table}

\begin{figure}
 \begin{center}
   \hspace*{-1cm}
   \includegraphics[width=10cm]{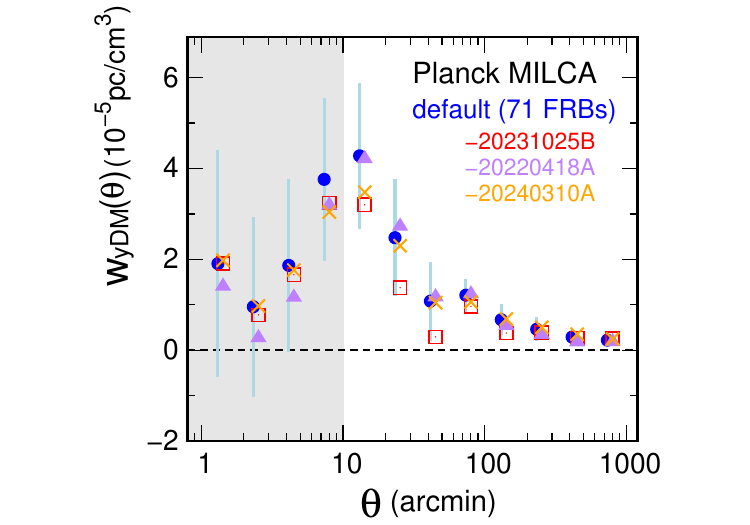}   %{fig_xi_topA-FRBs_paper.pdf} 
 \end{center}
\caption{Similar to Fig.~\ref{fig_xi_top-FRBs}, but excluding each of the top three FRBs that produce a large $\mathcal{A}$ (in Table \ref{table:Top5A}). The blue symbols with error bars represent our default. The red squares, purple triangles, and orange crosses show the results of excluding 20231025B, 20220418A, or 20240310A from the default, respectively. 
%{Alt text: The contribution from 20231025B is apparent at $\theta \simeq 10^\prime$--$50^\prime$.}
}
\label{fig_xi_topA-FRBs}
\end{figure}

This subsection identifies the FRBs contributing to the cross-correlation signal in Planck.
For this purpose, we include the cluster FRBs (20220914A, 20231206A, and 20231229A) to evaluate their contributions, which are excluded from our default analysis.
The cross-correlation for each FRB is calculated over two angular ranges: $\theta=1^\prime$--$10^\prime$ and  $\theta=10^\prime$--$100^\prime$.
The small-scale signal at $\theta < 10^\prime$ is sensitive to its local environment but is blurred by the detector’s beam size.
The top five contributors are listed in Table \ref{table:Top5}.
In general, these FRBs show significant positive or negative $\Delta {\rm DM}_{\rm ext}$ (i.e., outliers in the ${\rm DM}$--$z$ relation) and redshifts below the average (except for 20220224C and 20231025B).
Notably, the FRBs with negative $\Delta {\rm DM}_{\rm ext}$ exist in low $y$-value environments, yielding significant positive cross-correlations.  
The cluster FRBs 20220914A and 20231206A obtain strong correlations in $\theta=1^\prime$--$10^\prime$ and $10^\prime$--$100^\prime$, respectively.  
The host cluster of 20231206A is located nearby (at $z=0.038$) and occupies $\sim 3^\circ \times 3^\circ$ of the sky~\citep{Rines2000,Amiri2025}, thereby influencing the signal extending to large angular scales.
A foreground cluster at $z=0.0639$ contributes to the large ${\rm DM}_{\rm obs}$~\citep{Pastor2025} of 20220224C.
The host of 20240114A is a star-forming dwarf galaxy associated with a more massive central galaxy~\citep{Tian2024,Bhardwaj2025}. 
Its large $\Delta {\rm DM}_{\rm ext}$ comes from the central-galaxy halo, a foreground cluster at $z=0.09$, and eight foreground galaxies whose virial radii intersect with the source’s sight line~\citep{Bhardwaj2025}.

Figure \ref{fig_xi_top-FRBs} illustrates the contributions of the cluster FRBs.
These FRBs, especially 20220914A, significantly generate the signal at $\theta \lesssim 10^\prime$ because the small-scale signal is sensitive to the local environment and the line-of-sight foreground structures.
Furthermore, because the number of angular pairs between FRBs and the $y$-map is limited on such small scales, a few FRBs can greatly influence the cross-correlation.
In contrast, over large scales ($\theta \gtrsim 10^\prime$), the correlations are not dominated by a few specific FRBs but by many FRBs making approximately equal contributions.
The host-cluster contributions appear less significant in Fig.~\ref{fig_xi_top-FRBs} than in Fig. \ref{fig_xi_host-ave}, even at small scales, because the host masses are relatively small\footnote{The tSZ masses are $M_{500 {\rm c}}/(10^{14} M_\odot) = 1.94^{+0.28}_{-0.31}$, $2.13 \pm 0.15$, and $1.12 \pm 0.11$ for 20220914A, 20231206A, and 20231229A~\citep{BH2024}, respectively, converted to the virial mass $M_{\rm host} \approx 2 M_{500{\rm c}}$ (e.g., \cite{KS2001}).} ($M_{\rm host} \approx (1$--$3) \times 10^{14} h^{-1} M_\odot$).
In summary, Fig.~\ref{fig_xi_top-FRBs} indicates that the host contribution is negligible at $\theta \gtrsim 10^\prime$.

We also calculate each FRB's contribution to the amplitude $\mathcal{A}$. First, we compute $\mathcal{A}$ by combining the default $71$ FRBs with the three cluster FRBs. The result is $\mathcal{A} = 2.15 \pm 0.52$, close to the default $\mathcal{A} = 2.01 \pm 0.50$. This suggests that the cluster FRBs do not significantly change $\mathcal{A}$ because they mainly affect small-scale signals ($\theta \lesssim 10^\prime$), which are excluded from the analysis. Then, we remove each FRB from the $74$-source set and recompute $\mathcal{A}$. The top five contributors to $\mathcal{A}$ are listed in Table \ref{table:Top5A}. In general, these have large positive or negative $\Delta {\rm DM}_{\rm ext}$; in particular, three of them have negative $\Delta {\rm DM}_{\rm ext}$. Two of them (20231025B and 20240310A) also rank among the top five in the cross-correlation in Table \ref{table:Top5}. 20231025B contributes the most by far, accounting for approximately half of $\mathcal{A}$. Its contribution to the cross-correlation is significant at $\theta \simeq 10^\prime$--$50^\prime$, as plotted in Fig.~\ref{fig_xi_topA-FRBs}. Among the $74$ FRBs, it has the fourth-smallest $\Delta {\rm DM}_{\rm ext}$ and the second-lowest average $y$ (i.e., negative $\delta y$ in Eq.~\ref{eq:w_yDM_estimator}) within a circle of radius $100^\prime$ around it, which explains its large contribution. The negative $\Delta {\rm DM}_{\rm ext}$ and low $y$ indicate that its foreground is underdense in electrons.

We note that the cross-correlation $\hat{w}_{y{\rm DM}}$ and the amplitude $\mathcal{A}$ are different statistics; $\hat{w}_{y{\rm DM}}$ is the correlation over a given angular range (e.g., $10^\prime$--$100^\prime$), which is strongly influenced by the correlation near the outer angular radius (e.g., $100^\prime$) because there are more pairs at larger angles, whereas $\mathcal{A}$ measures the amplitude over wider angular scales ($10^\prime$--$1000^\prime$) with accounting for its covariance. Therefore, a correlation between them (such as a large/small $\hat{w}_{y{\rm DM}}$ leading to a large/small $\mathcal{A}$) may not hold exactly in general.

\subsection{Parameter dependence of the measurements}
\label{sec:w_param_depend}

\begin{table*}
 \tbl{Same as Table \ref{table_bestfit_DM-z} (the N$\beta$ model), but with $\sigma_8$ and $T_{\rm AGN}$ varied from the default values ($\sigma_8 = 0.811$ and $\log_{10}(T_{\rm AGN}/{\rm K}) = 7.8$). The last column gives the best-fit amplitude with a $68 \%$ confidence interval.}
{\begin{tabular}{cccccc}
      \hline
       Model & $f_{\rm e}$ & ${\rm DM}_{\rm host,0}$ (${\rm pc/cm}^3$) & $\sigma_{\rm host,0}$ (${\rm pc/cm}^3$) & $\beta_{\rm host}$ & $\mathcal{A}$ \\ 
      \hline
       default & $\mathbf{0.972}$ $(0.918^{+0.082}_{-0.023})$ & $\mathbf{120.2}$ $(130.6^{+15.3}_{-24.0})$ & $\mathbf{98.9}$ $(114.9^{+16.8}_{-41.0})$ & $\mathbf{-0.199}$ $(-0.327^{+0.776}_{-1.070})$ & $2.01 \pm 0.50$ \\
       \hdashline
       $\sigma_8$ $+ 5\%$ & $\mathbf{0.975}$ $(0.915^{+0.084}_{-0.024})$ & $\mathbf{118.6}$ $(128.7^{+14.9}_{-22.7})$ & $\mathbf{91.6}$ $(106.2^{+15.5}_{-36.9})$ & $\mathbf{-0.171}$ $(-0.371^{+0.767}_{-1.017})$ & $1.55 \pm 0.39$ \\
       $\sigma_8$ $- 5\%$ & $\mathbf{0.964}$ $(0.919^{+0.080}_{-0.022})$ & $\mathbf{121.7}$ $(132.5^{+15.0}_{-25.2})$ & $\mathbf{104.8}$ $(123.8^{+16.9}_{-45.6})$ & $\mathbf{-0.223}$ $(-0.273^{+0.819}_{-1.068})$ & $2.66 \pm 0.66$ \\
       $\log T_{\rm AGN}\!=\!7.6$ & $\mathbf{0.973}$ $(0.911^{+0.089}_{-0.027})$ & $\mathbf{114.5}$ $(125.3^{+14.4}_{-21.2})$ & $\mathbf{77.5}$ $(93.5^{+16.0}_{-32.0})$ & $\mathbf{-0.366}$ $(-0.510^{+0.712}_{-0.973})$ & $1.78 \pm 0.46$ \\
       $\log T_{\rm AGN}\!=\!8.0$ & $\mathbf{0.932}$ $(0.911^{+0.088}_{-0.027})$ & $\mathbf{122.8}$ $(135.6^{+15.2}_{-27.4})$ & $\mathbf{113.5}$ $(140.1^{+18.0}_{-54.3})$ & $\mathbf{-0.480}$ $(-0.311^{+0.804}_{-1.145})$ & $2.46 \pm 0.63$ \\
      \hline
    \end{tabular}}
    \label{table_param-depend}
\begin{tabnote}
%\footnotemark[]In the second and fourth rows, $\beta_{\rm host}$ is fixed at 1. 
%NE2001+YT20 is used for ${\rm DM}_{\rm MW}$ in the first two rows, while YMW16+YT20 is used in the last two rows.
\end{tabnote}
\end{table*}

\begin{figure}
 \begin{center}
   \hspace*{-1cm}
   \includegraphics[width=10cm]{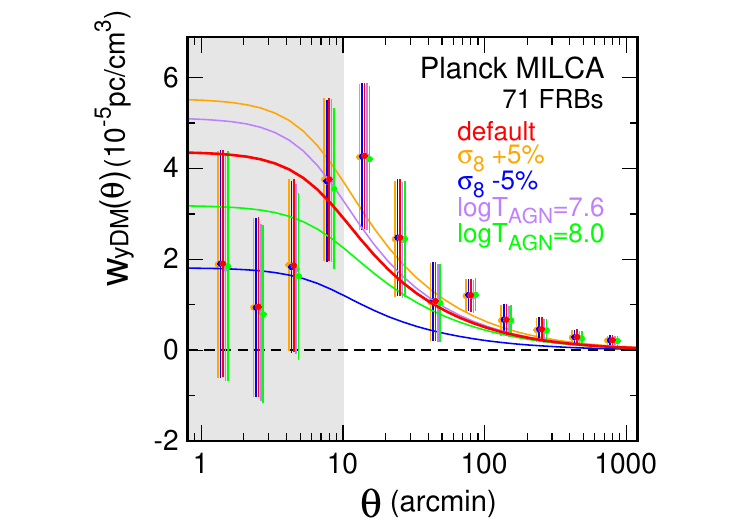}    %{fig_xi_Planck_param-depend_paper.pdf}
 \end{center}
\caption{Parameter dependencies of the cross-correlation measurements and theoretical predictions. The red symbols and curve use our default values ($\sigma_8 = 0.811$ and $\log_{10}(T_{\rm AGN}/{\rm K}) = 7.8$). The orange and blue ones vary $\sigma_8$ by $\pm 5 \%$, respectively. The purple and green ones change $\log_{10} (T_{\rm AGN}/{\rm K})$ to $7.6$ and $8.0$, respectively. 
%The measurements are also slightly influenced by these parameters through $\Delta {\rm DM}_{\rm  ext}$. 
The symbols are slightly offset along the $x$-axis.   
%{Alt text: Same as the left panel of Fig.~\ref{fig_xi_obs}, but for different values of $\sigma_8$ and $T_{\rm AGN}$.}
}
\label{fig_xi_Planck_param-depend}
\end{figure}

Because the theoretical cross-correlation is sensitive to $\sigma_8$ and $T_{\rm AGN}$ (as discussed in Subsubsection \ref{sec:theoretical_w_yDM}), this subsection examines their parameter dependence by varying $\sigma_8$ by $\pm 5 \%$ and $\log_{10}(T_{\rm AGN}/{\rm K})$ by $\pm 0.2$ relative to the default values ($\sigma_8 = 0.811$ and $\log_{10}(T_{\rm AGN}/{\rm K}) = 7.8$). First, we fit the model parameters for the ionized fraction ($f_{\rm e}$) and the host properties (${\rm DM}_{\rm host,0}$, $\sigma_{\rm host,0}$, and $\beta_{\rm host}$) using the ${\rm DM}$--$z$ analysis, following the same procedure as in Subsection \ref{sec:full-skymap_DeltaDMe}. The best-fit parameters are listed in Table \ref{table_param-depend}. The two parameters ($\sigma_8$ and $T_{\rm AGN}$) affect the variance of ${\rm DM}_{\rm cos}$, $\sigma_{\rm DM,cos}^2$, through the electron power spectrum in Eq.~(\ref{eq:sigma_DMcos}). For smaller (larger) $\sigma_8$ or larger (smaller) $T_{\rm AGN}$, $\sigma_{\rm DM,cos}$ decreases (increases), so $\sigma_{\rm host,0}$ increases (decreases) to compensate, as shown in Table \ref{table_param-depend}. The other parameters ($f_{\rm e}$, ${\rm DM}_{\rm host,0}$, and $\beta_{\rm host}$) are less sensitive to $\sigma_8$ and $T_{\rm AGN}$. Then, we calculate the residual ($\Delta {\rm DM}_{\rm ext}$) using Eq.~(\ref{eq:dDM_ext}) with these best-fit values and measure the cross-correlation following the same procedure as in Subsection \ref{sec:w_yDM_estimator}. 
%Therefore, we caution that the cross-correlation measurement depends on the given values of  $\sigma_8$ and $T_{\rm AGN}$. 
Figure \ref{fig_xi_Planck_param-depend} shows the measurement results, which clearly indicate that the measurements (shown as dots with error bars) are quite insensitive to $\sigma_8$ and $T_{\rm AGN}$. The solid curves show the theoretical prediction; $\sigma_8$ affects the results at all scales, whereas $T_{\rm AGN}$ mainly affects small scales ($\theta \lesssim$ a few$\times 10^\prime$). The constraint on the amplitude $\mathcal{A}$ is given in Table \ref{table_param-depend}; $\mathcal{A}$ is larger for smaller $\sigma_8$ or larger $T_{\rm AGN}$, and vice versa.

We have used two steps: the first constrains $f_{\rm e}$, ${\rm DM}_{\rm host,0}$, $\sigma_{\rm host,0}$, and $\beta_{\rm host}$ using the ${\rm DM}$--$z$ analysis, and the second constrains $\mathcal{A}$ using the cross-correlation result. In principle, combining these steps simultaneously allows us to constrain $\sigma_8$ and $T_{\rm AGN}$. However, we need to address the first and second steps iteratively many times, which is quite time-consuming. We will leave the constraints on $\sigma_8$ and $T_{\rm AGN}$ for future work.

\subsection{Constraints on the gas temperature}
\label{sec:constraint_Te}

\begin{figure*}[t]
        \begin{center}
        \includegraphics[width=16cm]{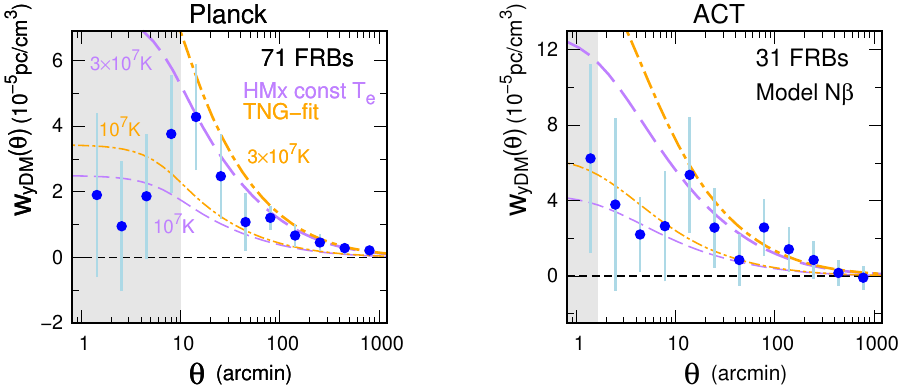}   %{fig_xi_Planck-ACT_constTe_paper.pdf}
        \end{center}
\caption{Same as Fig.~\ref{fig_xi_obs}, but compared with the constant gas temperature model. 
The dashed purple and dot-dashed orange curves represent the theoretical \texttt{HMx} and TNG-fit models, respectively, at $T_{\rm e} = 10^7 \, {\rm K}$ (thin) and $3 \times 10^7 \, {\rm K}$ (thick). 
%{Alt text: Line graphs and data points indicate the temperature is roughly twenty million kelvins.}
}
\label{fig_xi_Planck_constTe}
\end{figure*}

\begin{table}
\tbl{Constraints on the gas temperature based on the two theoretical \texttt{HMx} and TNG-fit models. The results are the means $\pm$ $68 \%$ credible intervals in units of $10^7 ~{\rm K}$.}
%The fitting range is $\theta=10^\prime$--$316^\prime$. %for Planck and $\theta=1^\prime.78$--$316^\prime$ for ACT.
%Model N$\beta$ is used.}
{%
\begin{tabular}{ccc}
      \hline
       &  Planck MILCA & ACT \\
      \hline
       \texttt{HMx} & $2.69 \pm 0.67$ & $2.25 \pm 1.32$ \\
       TNG-fit & $2.38 \pm 0.59$ & $1.71 \pm 1.05$ \\
       %\texttt{HMx} w/ Y$\beta$ & $1.80 \pm 0.61$ & $1.00 \pm 0.55$ & $0.98 \pm 0.86$ \\
       %TNG-fit w/ Y$\beta$ & $1.78 \pm 0.58$ & $0.99 \pm 0.53$ & $0.95 \pm 0.80$ \\
      \hline
    \end{tabular}}
    \label{table:Te}
\begin{tabnote}
\end{tabnote}
\end{table}

This subsection provides the constraints on the gas temperature $T_{\rm e}$ using the constant $T_{\rm e}$ model based on \texttt{HMx} and TNG-fit (Subsection \ref{sec:pk_constTe}). %as the theoretical correlation.
Since the cross-correlation is directly proportional to $T_{\rm e}$, the likelihood analysis of $T_{\rm e}$ is similar to that of the amplitude $\mathcal{A}$ in Subsection \ref{sec:w_yDM_results}. 
Table \ref{table:Te} shows the best-fit parameters for $T_{\rm e}$, indicating $T_{\rm e} \approx 2 \times 10^7 \, {\rm K}$. %though the confidence intervals are relatively large.
\texttt{HMx} obtains a higher temperature than TNG-fit because it provides stronger feedback and therefore predicts a lower cross-correlation for a given $T_{\rm e}$ (Fig.~\ref{fig_xi_constTe}). 
Figure \ref{fig_xi_Planck_constTe} shows the theoretical predictions at $T_{\rm e} = 10^7 \, {\rm K}$ and $3 \times 10^7 \, {\rm K}$, which fairly agree with the Planck and ACT measurements.

\citet{VanWaerbeke2014} previously constrained the temperature using an angular cross-correlation between the $y$-map of the Planck nominal data~\citep{Planck2014_overview} and the weak lensing mass map from the Canada-France-Hawaii Telescope Lensing Survey~\citep{VanWaerbeke2013}. %(CFHTLenS; \cite{VanWaerbeke2013}). 
They found a positive correlation at $\theta=0^\prime$--$60^\prime$ and constrained the temperature as $(b_{\rm e}/1) (T_{\rm e}/0.1 \, {\rm keV}) (\bar{n}_{\rm e}/1 \, {\rm m}^{-3}) \simeq 2.0$ at $z = 0$. 
Using the mean free-electron density in Eq.~(\ref{eq:ne_mean}), this constraint is rewritten as $T_{\rm e} \simeq 1.2 \times 10^7 \, {\rm K} \, (b_{\rm e}/1)^{-1} (f_{\rm e}/0.9)^{-1}$, consistent with our result.
\citet{Ibitoye2024} recently measured a cross-power spectrum between the integrated Sachs-Wolfe effect and the Planck $y$-map and provided a similar constraint: $T_{\rm e} \simeq 1.8 \times 10^7 \, {\rm K} \, (b_{\rm e}/1)^{-1} (f_{\rm e}/0.9)^{-1}$.

\citet{VanWaerbeke2014} concluded that the correlation signal comes from diffuse gas tracing the large-scale structure.
Accordingly, they attributed their measured temperature to this gas. 
However, \citet{Battaglia2015} later argued that the signal is primarily influenced by hot gas in ICM.
According to \texttt{HMx}, the correlation signal mainly originates from massive halos ($\gtrsim 10^{14} \, M_\odot/h$) with smaller contributions from diffuse gas (Subsection \ref{sec:HMx}).
Observations using the tSZ effect, kinetic SZ effect, and/or X-ray measurements have revealed gas at temperatures of $\sim 10^7 \, {\rm K}$ in the outskirts of clusters (e.g., \cite{Eckert2013,Ghirardini2019}), massive galaxy halos~\citep{Schaan2021}, and the filaments connecting galaxies~\citep{Tanimura2022}. 
Therefore, the measured temperature can be attributed to these structures.

\section{Discussion}    %Known issues and uncertainties}
\label{sec:discussion}

\begin{figure}
 \begin{center}
   \hspace*{-1cm}
   \includegraphics[width=10cm]{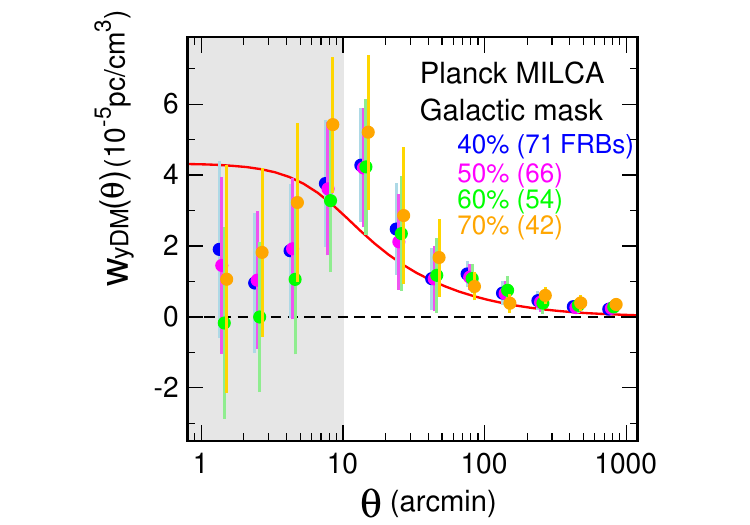}   %{fig_xi_Planck_mask_paper.pdf} 
 \end{center}
\caption{Effect of Galactic mask on the cross-correlation measurement. The masked region covers $40 \%$ to $70 \%$ of the sky along the Galactic plane. Our default setting is $40 \%$. The values in parentheses indicate the number of FRBs in the survey area. 
The symbols are slightly offset along the $x$-axis to prevent overlap.
The red curve represents the \texttt{HMx} prediction with the $40 \%$ mask (the same curve is shown in Fig.~\ref{fig_xi_obs}).
%{Alt text: Data points show that the choice of the Galactic mask does not impact the measurement.}
}
\label{fig_xi_Planck_mask}
\end{figure}

\begin{figure}
 \begin{center}
   \hspace*{-1cm}
   \includegraphics[width=10cm]{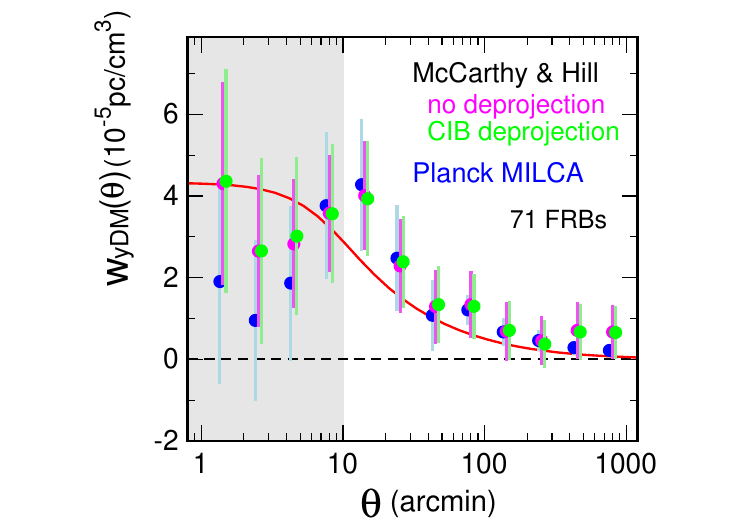}    %{fig_xi_Planck_CIB_paper.pdf} 
 \end{center}
\caption{Cross-correlation measurements using the McCarthy \& Hill $y$-maps constructed from the Planck PR4 data (magenta symbols) and the results after excluding CIB contamination (i.e., the deprojecting CIB; green symbols). The blue symbols are the cross-correlation measurements using the Planck MILCA map and the red curve shows the \texttt{HMx} prediction (both are also shown in Fig. \ref{fig_xi_obs}). The symbols are slightly offset along the $x$-axis.
%{Alt text: Data points indicate that the $y$-map selection does not affect the measurement.}
}
\label{fig_xi_Planck_CIB}
\end{figure}

This section discusses potential contamination from the Galactic foreground, the cosmic infrared background (CIB), and non-thermal gas pressure in the cross-correlation measurement. 
%including , as well as the cross-power spectrum analysis.

The tSZ signal is contaminated by thermal dust, primarily associated with the Galactic plane (e.g., \cite{Planck2016_ymap}). 
Furthermore, ${\rm DM}_{\rm MW}$ substantially contributes near the Galactic plane, implying that uncertainty in the models (NE2001, YMW16 and YT20) affects the ${\rm DM}_{\rm ext}$ estimation.  %in Eq.~(\ref{eq:DM_ext}). 
To examine contamination near the Galactic plane, we replace our default setting (the Galactic $40 \%$ mask) with the more conservative Galactic masks\footref{ftn_P} ranging from $50 \%$ to $70 \%$. 
Figure \ref{fig_xi_Planck_mask} shows the measurements with these masks. 
The results largely overlap, indicating non-significant contamination from the Galactic plane. 
The large scatter among the plots on small scales ($\theta < 10^\prime$) is caused by the limited number of pairs between $y$ and $\Delta {\rm DM}_{\rm ext}$. 
The large scatter with the $70 \%$ mask is explained by the small number of FRBs in the survey area.

A major contaminant of the tSZ effect is the CIB produced by thermal dust in distant galaxies at $z\simeq1$--$3$ (e.g., \cite{Mroczkowski2019}). 
The CIB affects the high-frequency maps ($\gtrsim 143$ GHz) and the resulting $y$-map~\citep{Planck2016XI}. 
It also contaminates the cross-correlation between a tracer and the $y$-map when the tracer is located within the redshift range of dusty galaxies. 
However, as the current FRBs (with a mean redshift of $0.26$) are much closer than dusty galaxies at $z\simeq1$--$3$, CIB will minimally contaminate the cross-correlation.
\citet{MH2024a} recently provide the $y$-maps\footnote{\url{https://users.flatironinstitute.org/~fmccarthy/ymaps_PR4_McCH23/}} from the Planck PR4 data~\citep{PlanckPR42020} using their NILC pipeline (see also \cite{MH2024b}). 
They also offer the CIB deprojected $y$-map, which removes the CIB contamination assuming its intensity is $I_{\rm CIB} \propto \nu^{\beta_{\rm CIB}} B(\nu;T_{\rm CIB})$ where $B(\nu;T)$ is the black-body spectrum at frequency $\nu$ and temperature $T$. 
Figure \ref{fig_xi_Planck_CIB} shows the cross-correlations obtained through the procedure described in Section \ref{sec:measurment}, but replacing the Planck MILCA map with the McCarthy and Hill maps (setting $\sigma_y=1$ in Eq.~(\ref{eq:weight_w_yDM})). 
The CIB deprojection results are obtained with their default CIB model parameters $\beta_{\rm CIB}=1.7$ and $T_{\rm CIB}=10.71 \, {\rm K}$ in \citet{MH2024a}. 
The deprojected and non-deprojected results are nearly identical, suggesting a low effect from CIB contamination. 
The non-deprojected McCarthy and Hill results lie within the error bars of our Planck MILCA results, further validating our measurement.

This study has ignored non-thermal gas pressure arising from bulk motion and turbulence in clusters. Non-thermal electrons contribute to ${\rm DM}$ (as thermal electrons do) but not to $y$. The fraction of non-thermal pressure relative to the total pressure increases with radius and may reach $10$--$30 \%$ at $r_{\rm vir}$ (e.g., \cite{Pratt2019}), thereby suppressing $y$ and $w_{y{\rm DM}}$. Recently, \citet{LaPosta2024} developed a halo model that includes non-thermal pressure and constrained it using cross-correlations among the tSZ, weak lensing, and X-ray signals. Future measurements of the $y$-${\rm DM}$ cross-correlation, combined with cross-correlations involving other probes, will improve constraints on non-thermal pressure.

A pipeline for measuring the angular power spectra of discrete sources, such as FRB ${\rm DMs}$, has been recently developed (e.g., \cite{Wolz2025}). 
It can be used to analyze the cross-power spectrum between sparse ${\rm DMs}$ and the continuous $y$-map.
This analysis will be explored in future work.

\section{Conclusion}
\label{sec:conclusion}

This paper investigated the angular cross-correlation between the cosmological ${\rm DM}$ and the Compton $y$ parameter. 
First, we developed the theoretical cross-correlation using the halo model \texttt{HMx} (Subsection \ref{sec:HMx}).
The cross-correlation signal is mainly contributed by intervening massive clusters with $M \gtrsim 10^{14} \, h^{-1} M_\odot$ (Fig.~\ref{fig_xi_param}).
Examining the dependencies of cross-correlation on the input parameters, we observed that it is most sensitive to $\sigma_8$, similar to the tSZ power spectrum (Subsection \ref{sec:theoretical_w_yDM}; Eq.~(\ref{eq:w_theo_param-depend})).
We further established that small-scale signal ($\theta \lesssim 30^\prime$) constrains the baryon feedback strength (Fig.~\ref{fig_xi_param}).
A simple phenomenological model assuming constant gas temperature is also presented (Subsection \ref{sec:pk_constTe}).
We then measured the cross-correlation over the range $\theta=1^\prime$--$1000^\prime$ using real data: the ${\rm DMs}$ obtained from $133$ localized FRBs and the $y$-maps taken from Planck and ACT. 
We divided the extragalactic ${\rm DM}$ into its mean and residual using the ${\rm DM}$--$z$ relation and cross-correlated the residual with $y$.
The measurement signal was consistent with the amplitudes of $\mathcal{A}=2.01 \pm 0.50$ and $1.23 \pm 0.82$ for Planck and ACT, where $\mathcal{A}=1$ corresponds to the \texttt{HMx} theoretical prediction in the Planck 2018 best-fit $\Lambda$CDM model (Subsection \ref{sec:w_yDM_results}; Table \ref{table:amp}). 
To our knowledge, this is the first detection (at $4.0 \sigma$ level) of the $y$-${\rm DM}_{\rm cos}$ cross-correlation. 
Based on the measured amplitude, we estimated the average gas temperature as $\approx 2 \times 10^7 \, {\rm K}$  (Subsection \ref{sec:constraint_Te}; Table \ref{table:Te}).
Approximately half of $\mathcal{A}$ for Planck originates from a single source, 20231025B (Subsection \ref{sec:w_yDM_topFRBs}). Somewhat counterintuitively (contrary to theoretical expectations), it has a negative $\Delta {\rm DM}_{\rm ext}$ in a low-$y$ environment, indicating that its foreground is underdense in electrons. Many more localized FRBs would be required to determine the overall trend among FRBs contributing to the cross-correlation.

Several systematic errors or contaminations are present in the measurement results. 
The main contamination source is the massive host's contribution to the cross-correlation, which dominates the small-scale signal at $\theta \lesssim 10^\prime$ (Section \ref{sec:host_galaxy_corr}; Fig.~\ref{fig_xi_host}). 
To mitigate this problem, we removed the cluster FRBs from our analysis; as a result, the contamination in the correlation signal was noticeably reduced (Subsection \ref{sec:w_yDM_topFRBs}; Fig.~\ref{fig_xi_top-FRBs}).
The Galactic foreground and CIB are also potential sources of contamination; however, when investigated, they negligibly affected the current measurements (Section \ref{sec:discussion}). 
Detailed studies on other systematic issues are left for future work.

\begin{ack}
We thank Tomoka Sakurai, Qiyun Huang, and Teppei Okumura for useful discussions.
We also thank all scientists who made their valuable observational data of FRBs and $y$-maps publicly available.
%Acknowledgment should be placed at end of main text.
%(NOT after the Appendix.)
\end{ack}

\section*{Funding}
 This research is supported in part by JSPS KAKENHI grant Nos. 22H00130 (RT and KI), 20H05855 (RT), 23H04900 (KI), 23H05430 (KI), 23H01172 (KI), 26H02045 (KI), 24H00215 (KO), 25K17380 (KO), 25H01513 (KO), and 25H00662 (KO).

\section*{Data availability} 
The measurement data will be shared on reasonable request to the first author.
% The data underlying this article are available ...  
% Sample Data Availability Statements 
% https://academic.oup.com/pages/open-research/research-data#Data%20Availability%20Statements

\appendix %%%%%%%%%%%%%%%%%%%%%%%%%%%%%%%%%%%%%%%%%%%%%%%%%%%%%%%%

\section{List of localized FRBs}
\label{app:list_localized_FRB}

Table \ref{table_list_FRBs1} presents our list of localized FRBs. 

\begin{table*}[t]
 \tbl{List of $133$ localized FRBs: Name, equatorial coordinates (RA,DEC), observed ${\rm DM}$, redshift, and reference. The term ``P/A'' in the $y$-map column indicates that the FRB is located within the survey area of Planck/ACT.}{%
 \begin{tabular}{lllcccc}
      \hline
       name &  RA   & DEC   &  ${\rm DM}_{\rm obs}$ & redshift & $y$-map & reference  \\ 
       &  (deg)  & (deg)  & (${\rm pc/cm}^3$) &  & &   \\ 
      \hline
        20121102A  & 82.995 & 33.148 & 557 & 0.1927 & & \citet{Chatterjee2017} \\
        20150418A & 109.129 & -19.040 & 776.2 & 0.492 & & \citet{Keane2016} \\
        20171020A  & 333.853 & -19.585 & 114.1 & 0.0087 & PA & \citet{Mahony2018,Lee-Waddell2023}  \\
        20180301A  & 93.227 & 4.671 & 536 & 0.3305 & & \citet{Bhandari2022} \\
        20180814A & 65.683 & 73.664 & 189.4 & 0.06835 & & \citet{Michilli2023} \\
        20180916B  & 29.503 & 65.717 & 348.8 & 0.0337 & & \citet{Marcote2020} \\
        20180924B  & 326.105 & -40.900 & 362.16 & 0.3214 & PA & \citet{Bannister2019} \\
        20181030A  & 158.584 & 73.751 & 103.5 & 0.0039 & & \citet{Bhardwaj2021} \\
        20181112A  & 327.348 & -52.971 & 589 & 0.4755 & PA & \citet{Prochaska2019} \\
        20181220A & 348.698 & 48.342 & 209.4 & 0.02746 & & \citet{Bhardwaj2023} \\
        20181223C & 180.921 & 27.548 & 112.5 & 0.03024 & P & \citet{Bhardwaj2023} \\
        20190102C  & 322.416 & -79.476 & 364.55 & 0.2913 & & \citet{Macquart2020} \\
        20190110C & 249.318 & 41.443 & 221.6 & 0.12244 & P & \citet{Ibik2024} \\
        20190303A & 207.996 & 48.121 & 222.4 & 0.064 & P & \citet{Michilli2023} \\
        20190418A & 65.812 & 16.074 & 184.5 & 0.07132 & & \citet{Bhardwaj2023} \\
        20190425A & 255.663 & 21.577 & 128.2 & 0.03122 & P & \citet{Bhardwaj2023} \\
        20190520B & 240.518 & -11.288 & 1204.7 & 0.241 & & \citet{Niu2022}  \\
        20190523A  & 207.065 & 72.470 & 760.8 & 0.66 & P & \citet{Ravi2019} \\
        20190608B  & 334.020 & -7.898 & 340.05 & 0.1178 & PA & \citet{Macquart2020} \\
        20190611B  & 320.745 & -79.398 & 321.4 & 0.3778 & & \citet{Macquart2020} \\
        20190614D & 65.0755 & 73.707 & 959.2 & 0.6 & & \citet{Law2020} \\
        20190711A  & 329.420 & -80.358 & 592.6 & 0.5217 & & \citet{Macquart2020} \\
        20190714A  & 183.980 & -13.021 & 504.13 & 0.2365 & & \citet{Heintz2020} \\
        20191001A  & 323.352 & -54.748 & 507.9 & 0.234 & PA & \citet{Heintz2020} \\
        20191106C & 199.580 & 43.000 & 332.2 & 0.10775 & P & \citet{Ibik2024} \\
        20191228A  & 344.430 & -29.594 & 297.5 & 0.2432 & A & \citet{Bhandari2022}   \\
        20200223B & 8.270 & 28.831 & 201.8 & 0.0602 & P & \citet{Ibik2024} \\
        20200430A  & 229.706 & 12.377 & 380.25 & 0.1608 & PA & \citet{Heintz2020} \\
        20200723B & 190.158 & -5.135 & 244.05 & 0.0085 & & \citet{Shin2024} \\
        20200906A  & 53.499 & -14.083 & 577.8 & 0.3688 & PA & \citet{Bhandari2022} \\
        20201123A & 263.67 & -50.76 & 433.55 & 0.0507 & & \citet{Rajwade2022} \\
        20201124A  & 77.015 & 26.061 & 413.52 & 0.0979 & & \citet{Fong2021} \\
        20210117A & 339.979 & -16.152 & 729.1 & 0.214 & PA & \citet{Bhandari2023} \\
        20210320C & 204.458 & -16.123 & 384.8 & 0.2797 & P & \citet{James2022,Gordon2023} \\
        20210405I & 255.339 & -49.545 & 566.43 & 0.066 & & \citet{Driessen2024} \\
        20210410D & 326.086 & -79.318 & 578.78 & 0.1415 & & \citet{Caleb2023,Gordon2023}  \\
        20210603A & 10.274 & 21.226 & 500.15 & 0.1772 & P & \citet{Cassanelli2024}  \\
        20210807D & 299.221 & -0.762 & 251.9 & 0.1293 & & \citet{James2022,Gordon2023}  \\
        20211127I & 199.808 & -18.838 & 234.83 & 0.0469 & P & \citet{James2022,Gordon2023}  \\
        20211203C & 204.563 & -31.380 & 636.2 & 0.3439 & P & \citet{James2022,Gordon2023} \\
        20211212A & 157.351 & 1.361 & 206 & 0.0707 & PA & \citet{James2022,Gordon2023}  \\
        20220105A & 208.803 & 22.467 & 583 & 0.2785 & P & \citet{Gordon2023}  \\
        20220204A & 274.226 & 69.723 & 612.2 & 0.4 & P & \citet{Connor2024,Sharma2024}  \\
        20220207C  & 310.200 & 72.882 & 262.38 & 0.04304 & & \citet{Law2024} \\
        20220208A & 322.575 & 70.041 & 437 & 0.351 & & \citet{Connor2024,Sharma2024}  \\
        20220222C &	203.905 & -28.027 &	1071.2 & 0.853 & P & \citet{Pastor2025}  \\
        20220224C &	166.678 & -22.940 &	1140.2 & 0.6271 & P & \citet{Pastor2025}  \\
        20220307B  & 350.875 & 72.192 & 499.27 & 0.248123 & & \citet{Law2024} \\
        20220310F  & 134.72 & 73.491 & 462.24 & 0.477958 & P & \citet{Law2024} \\
        20220319D  & 32.178 & 71.035 & 110.98 & 0.011228 & & \citet{Law2024} \\
        20220330D & 163.751 & 70.351 & 468.1 & 0.3714 & & \citet{Connor2024,Sharma2024} \\
        20220418A  & 219.105 & 70.096 & 623.25 & 0.622 & P & \citet{Law2024} \\
        20220501C & 352.379 & -32.491 & 449.5 & 0.381 & PA & \citet{Shannon2025} \\
        20220506D  & 318.044 & 72.827 & 396.97 & 0.30039 & & \citet{Law2024}  \\
        20220509G  & 282.67 & 70.244 & 269.53 & 0.0894 & P & \citet{Law2024} \\
        20220529A & 19.104 & 20.632 & 250.2 & 0.1839 & P & \citet{Li2025} \\
        20220610A & 351.073 & -33.514 & 1458.15 & 1.016 & PA & \citet{Ryder2023}  \\
        20220717A & 293.304 & -19.288 & 637.34 & 0.363 & A & \citet{Rajwade2024} \\
        20220725A & 353.315 & -35.990 & 290.4 & 0.1926 & PA & \citet{Shannon2025} \\         
        20220726A & 73.946 & 69.930 & 686.55 & 0.361 & & \citet{Connor2024,Sharma2024} \\
        20220825A & 311.981 & 72.585 & 651.24 & 0.241397 & & \citet{Law2024} \\
        20220831A & 338.696 & 70.539 & 1146.25 & 0.262 & & \citet{Connor2024,Sharma2024} \\
        20220912A & 347.27 & 48.707 & 219.46 & 0.0771 & & \citet{Ravi2023,Zhang2023b} \\
        20220914A$^\dag$ & 282.056 & 73.337 & 631.28 & 0.1139 & P &\citet{Law2024} \\
        20220918A & 17.592 & -70.811 & 656.8 & 0.491 & P & \citet{Shannon2025} \\
        20220920A & 240.257 & 70.919 & 314.99 & 0.158239 & P & \citet{Law2024} \\
      \hline
    \end{tabular}}
    \label{table_list_FRBs1}
\begin{tabnote}
\end{tabnote}
\end{table*}

\setcounter{table}{7}

\begin{table*}[!t]
 \tbl{continued}{  %from Table \ref{table_list_FRBs1}}{
 %List of localized FRBs: its name, equatorial coordinates (RA,DEC), observed DM, redshift, and reference.}{%
 \begin{tabular}{lllcccc}
      \hline
      name &  RA   & DEC   &  ${\rm DM}_{\rm obs}$ & redshift & $y$-map & reference  \\ 
       &  (deg)  & (deg)  & (${\rm pc/cm}^3$) &  & &   \\ 
      \hline       
        20221012A & 280.798 & 70.524 & 441.08 & 0.284669 & P & \citet{Law2024} \\
        20221022A &	48.629 & 86.872 & 116.84 & 0.0149 & & \citet{Mckinven2025} \\
        20221027A & 130.872 & 72.101 & 452.5 & 0.229 & P & \citet{Connor2024,Sharma2024} \\
        20221029A & 141.964 & 72.453 & 1391.05 & 0.975 & & \citet{Connor2024,Sharma2024} \\
        20221101B & 342.216 & 70.682 & 490.7 & 0.2395 & & \citet{Connor2024,Sharma2024} \\
        20221106A & 56.705 & -25.570 & 343.8 & 0.2044 & A & \citet{Shannon2025} \\
        20221113A & 71.411 & 70.307 & 411.4 & 0.2505 & & \citet{Connor2024,Sharma2024} \\
        20221116A & 21.211 & 72.654 & 640.6 & 0.2764 & & \citet{Connor2024,Sharma2024} \\
        20221219A & 257.630 & 71.627 & 706.7 & 0.554 & P & \citet{Connor2024,Sharma2024} \\
        20230124A & 231.917 & 70.968 & 590.6 & 0.094 & & \citet{Connor2024,Sharma2024} \\
        20230125D &	150.205 & -31.545 &	640.1 & 0.3265 & & \citet{Pastor2025} \\
        20230203A & 151.662 & 35.694 & 420.1 & 0.1464 & P & \citet{Amiri2025} \\
        20230216A & 156.472 & 3.437 & 828 & 0.531 & PA & \citet{Connor2024,Sharma2024} \\
        20230222A & 106.960 & 11.225 & 706.1 & 0.1223 & P & \citet{Amiri2025} \\
        20230222B & 238.739 & 30.899 & 187.8 & 0.11 & P & \citet{Amiri2025} \\
        20230307A & 177.782 & 71.695 & 608.9 & 0.271 & P & \citet{Connor2024,Sharma2024} \\
        20230311A &	91.110 & 55.946 & 364.3 & 0.1918 & & \citet{Amiri2025} \\
        20230501A & 340.027 & 70.922 & 532.5 & 0.301 & & \citet{Connor2024,Sharma2024} \\
        20230506C & 12.100 & 42.006 & 772 & 0.3896 & & \citet{Anna-Thomas2025} \\
        20230521B & 351.036 & 71.138 & 1342.9 & 1.354 & & \citet{Connor2024,Sharma2024} \\
        20230526A & 22.233 & -52.717 & 361.4 & 0.157 & PA & \citet{Shannon2025} \\
        20230613A &	356.853 & -27.053 &	483.5 &	0.3923 & PA & \citet{Pastor2025}  \\
        20230626A & 235.630 & 71.134 & 451.2 & 0.327 & P & \citet{Connor2024,Sharma2024} \\
        20230628A & 166.787 & 72.282 & 345.15 & 0.1265 & P & \citet{Connor2024,Sharma2024} \\
        20230703A & 184.624 & 48.730 & 291.3 & 0.1184 & P & \citet{Amiri2025} \\
        20230708A & 303.115 & -55.356 & 411.51 & 0.105 & A & \citet{Shannon2025} \\
        20230712A & 167.359 & 72.558 & 586.96 & 0.4525 & P & \citet{Connor2024,Sharma2024}  \\
        20230718A & 128.162 & -40.452 & 477 & 0.035 & & \citet{Glowacki2024}  \\
        20230730A & 54.665 & 33.159 & 312.5 & 0.2115 & & \citet{Amiri2025} \\
        20230808F & 53.304 & -51.935 & 653.2 & 0.3472 & PA & \citet{Hanmer2025} \\
        20230814B & 335.975 & 73.026 & 696.4 & 0.5535 & & \citet{Connor2024,Sharma2024} \\
        20230902A & 52.140 & -47.334 & 440.1 & 0.3619 & PA & \citet{Shannon2025} \\
        20230907D &	187.143 & 8.658 & 1030.8 & 0.4638 & A & \citet{Pastor2025} \\
        20230926A & 269.125 & 41.814 & 222.8 & 0.0553 & P & \citet{Amiri2025} \\
        20230930A & 10.507 & 41.417 & 456 & 0.0925 & & \citet{Anna-Thomas2025} \\
        20231005A & 246.028 & 35.449 & 189.4 & 0.0713 & P & \citet{Amiri2025} \\
        20231011A & 18.241 & 41.749 & 186.3 & 0.0783 & P & \citet{Amiri2025} \\
        20231017A & 346.754 & 36.653 & 344.2 & 0.245 & & \citet{Amiri2025} \\
        20231020B &	57.278 & -37.770 & 952.2 & 0.4775 & PA & \citet{Pastor2025} \\
        20231025B & 270.788 & 63.989 & 368.7 & 0.3238 & P & \citet{Amiri2025} \\
        20231120A & 143.984 & 73.285 & 438.9 & 0.0368 & & \citet{Connor2024,Sharma2024} \\
        20231123A & 82.623 & 4.476 & 302.1 & 0.0729 & & \citet{Amiri2025} \\
        20231123B & 242.538 & 70.785 & 396.7 & 0.2625 & P & \citet{Connor2024,Sharma2024} \\
        20231128A & 199.578 & 42.993 & 331.6 & 0.1079 & P & \citet{Amiri2025} \\
        20231201A & 54.589 & 26.818 & 169.4 & 0.1119 & & \citet{Amiri2025} \\
        20231204A & 207.999 & 48.116 & 221 & 0.0644 & P & \citet{Amiri2025} \\
        20231206A$^\dag$ & 112.443 & 56.256 & 457.7 & 0.0659 & P & \citet{Amiri2025} \\
        20231220A & 123.909 & 73.660 & 491.2 & 0.3355 & & \citet{Connor2024,Sharma2024} \\
        20231223C & 259.545 & 29.498 & 165.8 & 0.1059 & P & \citet{Amiri2025} \\
        20231226A & 155.364 & 6.110 & 329.9 & 0.1569 & PA & \citet{Shannon2025} \\
        20231229A$^\dag$ & 26.468 & 35.113 & 198.5 & 0.019 & P & \citet{Amiri2025} \\
        20231230A & 72.798 & 2.394 & 131.4 & 0.0298 & PA & \citet{Amiri2025} \\
        20240114A & 321.916 & 4.329 & 527.65 & 0.13 & PA & \citet{Tian2024} \\
        20240119A & 224.467 & 71.612 & 483.1 & 0.37 & & \citet{Connor2024,Sharma2024} \\
        20240123A & 68.263 & 71.945 & 1462 & 0.968 & & \citet{Connor2024,Sharma2024} \\
        20240201A & 149.906 & 14.088 & 374.5 & 0.042729 & A & \citet{Shannon2025} \\
        20240209A & 289.85 & 86.060& 176.57 & 0.1384 & & \citet{Eftenkhari2025} \\
        20240210A & 8.780 & -28.271 & 283.73 & 0.023686 & PA & \citet{Shannon2025} \\
        20240213A & 166.168 & 74.075 & 357.4 & 0.1185 & P & \citet{Connor2024,Sharma2024} \\
        20240215A & 268.441 & 70.232 & 549.5 & 0.21 & P & \citet{Connor2024,Sharma2024} \\
        20240229A & 169.984 & 70.676 & 491.15 & 0.287 & P & \citet{Connor2024,Sharma2024} \\
        20240304A &	136.331 & -16.167 &	652.6 &	0.2423 & P & \citet{Gordon2025,Shannon2025} \\
        20240304B &	182.997 & 11.813 & 2458.2 & 2.148 & PA & \citet{Caleb2025} \\
        20240310A & 17.622 & -44.439 & 601.8 & 0.127 & PA & \citet{Shannon2025} \\
        20240318A &	150.393 & 37.616 & 256.4 & 0.112 & P & \citet{Gordon2025,Shannon2025} \\
        20241228A &	216.386 & 12.025 & 246.53 & 0.1614 & PA & \citet{Curtin2025} \\
        20250316A &	182.435 & 58.849 & 161.82 & 0.0065 & P & \citet{CHIMEFRB2025} \\
        \hline
    \end{tabular}}
    \label{table_list_FRBs2}
\begin{tabnote}
%\footnotemark[The top two rows use YMW16+YT20 for ${\rm DM}_{\rm MW}$, while the bottom two rows use NE2001+YT20. In the second and fourth rows, $\beta_{\rm host}$ is fixed at 1.] 
\footnotemark[$\dag$] The cluster FRBs are excluded from the cross-correlation analysis.
\end{tabnote}
\end{table*}

\section{Normalization of the \texttt{HMx} power spectra}
\label{app:pk_norm}

Our electron-density contrast $\delta_{\rm e}$ is related to the \texttt{HMx} gas-density contrast $\delta^{\rm HMx}_{\rm gas}$ ($\equiv \delta \rho_{\rm gas}/\bar{\rho}_{\rm m}$) as $\delta_{\rm e}(\bfchi;z)=(\bar{\rho}_{\rm m}/\bar{\rho}_{\rm gas}(z)) \, \delta_{\rm gas}^{\rm HMx}(\bfchi;z)$ $=(\Omega_{\rm m}/\Omega_{\rm b}) \, [1-(\Omega_{\rm m}/\Omega_{\rm b})(\bar{\rho}_{\rm star}(z)/\bar{\rho}_{\rm m})]^{-1} \delta_{\rm gas}^{\rm HMx}(\bfchi;z)$, where $\bar{\rho}_{\rm m}, \bar{\rho}_{\rm gas}$ and $\bar{\rho}_{\rm star}$ are the mean comoving densities of matter, gas, and stars, respectively ($\bar{\rho}_{\rm m}$ is constant while the others are functions of $z$). The ratio $\bar{\rho}_{\rm star}(z)/\bar{\rho}_{\rm m}$ is obtained by averaging the stellar fraction (Eq. (27) in \cite{Mead2020}) over all halo masses. 
In summary, our free-electron power spectrum and the \texttt{HMx} gas power spectrum are related as follows:
\begin{equation}
   P_{n_{\rm e}}(k;z) = \left( \frac{\Omega_{\rm m}}{\Omega_{\rm b}} \right)^{2} \left( 1- \frac{\Omega_{\rm m}}{\Omega_{\rm b}} \frac{\bar{\rho}_{\rm star}(z)}{\bar{\rho}_{\rm m}} \right)^{-2} P_{\rm gas}^{\rm HMx}(k;z).
\end{equation}

Similarly, since the ionized fraction in \texttt{HMx} is $\bar{\rho}_{\rm gas}/\bar{\rho}_{\rm b}$ (where $\bar{\rho}_{\rm b}$ is the mean baryon density), its pressure perturbation $\delta p^{\rm HMx}_{\rm e}$ should scale proportionally to the ionized fraction as $\delta p_{\rm e}(\bfchi;z)=f_{\rm e} \, (\bar{\rho}_{\rm b}/\bar{\rho}_{\rm gas}) \delta p_{\rm e}^{\rm HMx}(\bfchi;z)$ $=f_{\rm e} \, [1-(\Omega_{\rm m}/\Omega_{\rm b})(\bar{\rho}_{\rm star}(z)/\bar{\rho}_{\rm m})]^{-1} \delta p_{\rm e}^{\rm HMx}(\bfchi;z)$.
Therefore, our $P_{n_{\rm e}p_{\rm e}}(k;z)$ is related to the \texttt{HMx} gas-pressure power spectrum $P_{{\rm gas},p_{\rm e}}^{\rm HMx}$ as follows:
\begin{equation}
    P_{n_{\rm e}p_{\rm e}}(k;z) = f_{\rm e} \frac{\Omega_{\rm m}}{\Omega_{\rm b}} \left( 1- \frac{\Omega_{\rm m}}{\Omega_{\rm b}} \frac{\bar{\rho}_{\rm star}(z)}{\bar{\rho}_{\rm m}} \right)^{-2} P_{{\rm gas},p_{\rm e}}^{\rm HMx}(k;z).
\end{equation}
The cross-power spectrum between the matter density contrast and the pressure perturbation is given by
\begin{equation}
    P_{{\rm m},p_{\rm e}}(k;z) = f_{\rm e} \left( 1- \frac{\Omega_{\rm m}}{\Omega_{\rm b}} \frac{\bar{\rho}_{\rm star}(z)}{\bar{\rho}_{\rm m}} \right)^{-1} P_{{\rm m},p_{\rm e}}^{\rm HMx}(k;z).
\end{equation}

\section{Plot of the ${\rm DM}$--$z$ relation on a linear scale}
\label{sec:DM-z_linear}
\begin{figure}
 \begin{center}
   \includegraphics[width=8.5cm]{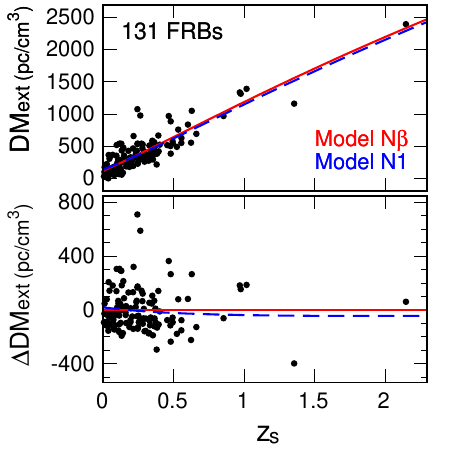}   %{fig_DM-z_linear_paper.pdf} 
 \end{center}
\caption{Same as Fig.~\ref{fig_DM-z}, but using a linear-linear scale. 
%{Alt text: Line graphs.}
}
\label{fig_DM-z_linear}
\end{figure}

Figure \ref{fig_DM-z_linear} is the same as Fig.~\ref{fig_DM-z} but uses a linear-linear scale.

\section{Off-diagonal elements of the covariance}
\label{app:cov}
\begin{figure}
 \begin{center}
   \includegraphics[width=10cm]{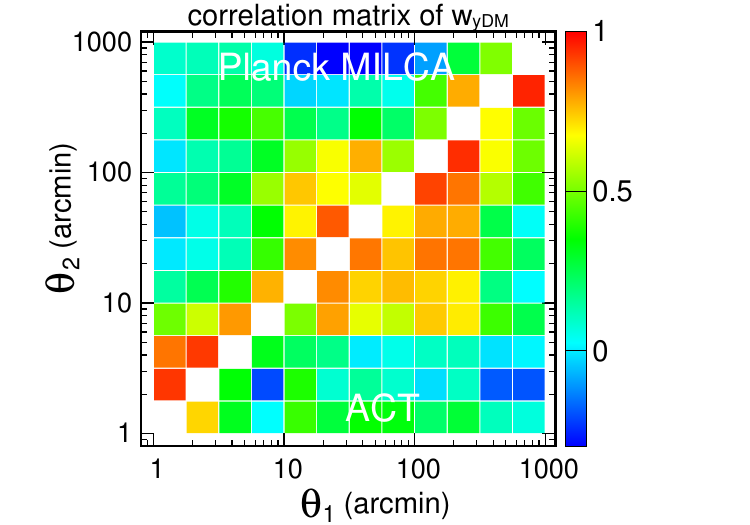}   %{fig_xi-cov_Planck-ACT_paper.pdf} 
 \end{center}
\caption{Off-diagonal elements of the correlation matrix. The upper-left and lower-right triangles are the results for Planck and ACT, respectively. 
%{Alt text: Color contour map.}
}
\label{fig_xi-cov}
\end{figure}

The correlation matrix of $w_{y{\rm DM}}(\theta)$ is defined in terms of its covariance as ${\rm Cov}(\theta_1,\theta_2)/\sqrt{{\rm Cov}(\theta_1,\theta_1) {\rm Cov}(\theta_2,\theta_2)}$. 
The off-diagonal elements, ranging from $-1$ to $1$, represent the correlation strengths between different angles $\theta_1$ and $\theta_2$. 
All diagonal elements are one. 
Figure \ref{fig_xi-cov} plots the off-diagonal elements for Planck MILCA with 71 FRBs and ACT with 31 FRBs using the N$\beta$ model. 
Positive correlations are observed, particularly among close angles.

\section{Consistency of cross-correlation measurements between Planck and ACT}
\label{app:consistency}

\begin{figure}
 \begin{center}
   \hspace*{-1cm}
   \includegraphics[width=10cm]{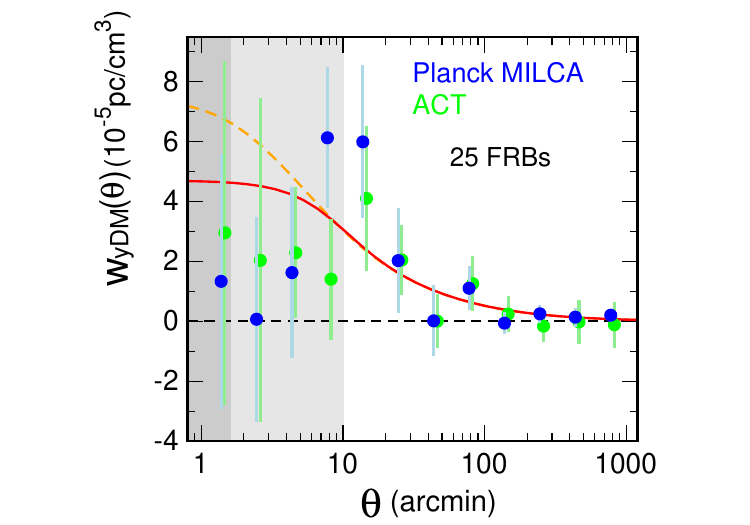}   %{fig_xi_Planck-ACT_overlap_paper.pdf}
 \end{center}
\caption{Cross-correlation measurement using $25$ FRBs that reside within both detectors' survey areas.
The blue (green) symbols show the results for Planck (ACT).
The solid red (dashed orange) curve is the theoretical prediction for Planck (ACT); the difference at $\theta \lesssim 10^\prime$ is due to the detectors’ different beam sizes.
%{Alt text: Measurement results using 25 sources that reside within both detectors’ survey areas.}
}
\label{fig_xi_PA_overlap}
\end{figure}

This section examines the consistency of cross-correlation results from Planck and ACT. To compare cross-correlations under the same conditions, we use $25$ FRBs that lie within both detectors’ survey regions. We also use the same $y$-map region (i.e., the overlapping $y$-map region of both detectors) and the same noise variance $\sigma_y^2 = 1$ in Eq.~(\ref{eq:weight_w_yDM}). Figure \ref{fig_xi_PA_overlap} shows the measurement results; they are consistent except for two data points near $\theta = 10^\prime$ (at $\theta = 8^\prime$ and $14^\prime$), where ACT yields a slightly lower amplitude. ACT shows a higher amplitude at $\theta < 5^\prime$, likely due to its better angular resolution. The best-fit amplitudes are $\mathcal{A} = 1.29 \pm 0.65$ for Planck and $\mathcal{A} = 1.49 \pm 1.02$ ($0.92 \pm 0.93$) for ACT, with a fitting range of $\theta = 10^\prime$--$1000^\prime$ ($1^\prime.78$--$1000^\prime$). Therefore, the results are consistent, though the constraint for ACT is somewhat weaker.
%ACT gives a lower $\mathcal{A}$ because 1) it has a lower correlation near $\theta = 10^\prime$, 
%2) it includes small-scale data at $\theta < 10^\prime$, 
%and 2) its theoretical prediction is higher at $\theta \lesssim 10^\prime$.}

%%%% 

\bibliographystyle{apj}
\bibliography{refs}

% Any journal's BST file (e.g., apj.bst) can be used as PASJ's BST is unavailable.    
% \bibliographystyle{****}
% \bibliography{****}
%\begin{thebibliography}{}
% Journals(e.g. A\&A,ApJ,AJ,NMRAS,PASP ...)
% Authors, Year, Journal, Vol#, Page# 
%\bibitem[Aauthor et al.(2001)]{key-1}
%  Aauthor, A., Bauthor, B., \& Cauthor, C.\ 2001, PASJ, 53, 000 
%\end{thebibliography}

\end{document}